%% file: LHCHiggsBRNote.tex
\documentclass[prd,amsmath,amssymb,floatfix,nofootinbib]{revtex4}

\usepackage{graphicx}% Include figure files
\usepackage{dcolumn}% Align table columns on decimal point
\usepackage{bm}% bold math
\usepackage{amsmath}
\newcommand{\met}{\rlap{\kern0.25em/}E_T}

\providecommand{\UGeV}{\unskip\,\mathrm{GeV}}
\providecommand{\UMeV}{\unskip\,\mathrm{MeV}}
\providecommand{\UTeV}{\unskip\,\mathrm{TeV}}

\newcommand{\Pe}{\mathrm{e}}
\newcommand{\Pf}{\mathrm{f}}
\newcommand{\Pfb}{\bar{\mathrm{f}}}
\newcommand{\Pl}{\mathrm{l}}
\newcommand{\Pq}{\mathrm{q}}
\newcommand{\Pqb}{\bar{\mathrm{q}}}
\newcommand{\PH}{\mathrm{H}}
\newcommand{\PV}{\mathrm{V}}
\newcommand{\PQu}{\mathrm{u}}
\newcommand{\PQd}{\mathrm{d}}
\newcommand{\PQs}{\mathrm{s}}
\newcommand{\PQc}{\mathrm{c}}
\newcommand{\PQb}{\mathrm{b}}
\newcommand{\PQt}{\mathrm{t}}
\newcommand{\PZ}{\mathrm{Z}}
\newcommand{\PW}{\mathrm{W}}
\newcommand{\Pg}{\mathrm{g}}
\newcommand{\PGg}{\gamma}
\newcommand{\alphas}{\alpha_\mathrm{s}}

\newcommand{\Ms}{m_\mathrm{s}}
\newcommand{\Mc}{m_\mathrm{c}}
\newcommand{\Mb}{m_\mathrm{b}}
\newcommand{\Mt}{m_\mathrm{t}}
\newcommand{\MZ}{M_\mathrm{Z}}
\newcommand{\MW}{M_\mathrm{W}}
\newcommand{\GF}{G_\mathrm{F}}

\newcommand{\MH}{M_\mathrm{H}}
\newcommand{\Hbb}{\PH \to \PQb \bar{\PQb}}
\newcommand{\Htautau}{\PH \to \tau^+\tau^-}
\newcommand{\Hmumu}{\PH \to \mu^+\mu^-}
\newcommand{\Hss}{\PH \to \PQs \bar{\PQs}}
\newcommand{\Hcc}{\PH \to \PQc \bar{\PQc}}
\newcommand{\Htt}{\PH \to \PQt \bar{\PQt}}
\newcommand{\Hgg}{\PH \to \Pg\Pg}
\newcommand{\Hgaga}{\PH \to \PGg\PGg}
\newcommand{\HZga}{\PH \to \PZ\gamma}
\newcommand{\HWW}{\PH \to \PW\PW}
\newcommand{\HZZ}{\PH \to \PZ\PZ}

\newcommand{\br}{{\mathrm{BR}}}
\newcommand{\Gatot}{\Gamma_{\mathrm H}}

\newcommand{\citere}[1]{\Bref{#1}}
\newcommand{\citeres}[1]{\Brefs{#1}}
\providecommand{\Bref}[1]{Ref.~\cite{#1}}
\providecommand{\Brefs}[1]{Refs.~\cite{#1}}
\newcommand{\refT}[1]{Table~\ref{#1}}
\newcommand{\refTs}[1]{Tables~\ref{#1}}
\newcommand{\refS}[1]{Section~\ref{#1}}

\newcommand{\refF}[1]{Figure~\ref{#1}}
\newcommand{\refFs}[1]{Figures~\ref{#1}}
\newcommand{\refE}[1]{(\ref{#1})}
\providecommand{\lsim}
{\;\raisebox{-.3em}{$\stackrel{\displaystyle <}{\sim}$}\;}
\providecommand{\gsim}
{\;\raisebox{-.3em}{$\stackrel{\displaystyle >}{\sim}$}\;}

\providecommand{\HDECAY}{{\sc HDECAY}}

\providecommand{\Prophecy}{{\sc Prophecy4f}}

\newcommand{\MSbar}{\ensuremath{\overline{\mathrm{MS}}}}

\begin{document}

\title{\vspace*{2.0cm}\Large LHC Higgs Cross Section Working Group \\  
\vspace*{0.7cm}
\huge Standard Model Higgs-Boson Branching Ratios with Uncertainties
\vspace*{0.7cm}
}

\author{A. Denner$^{1}$}
\author{S. Heinemeyer$^{2}$}
\author{I. Puljak$^{3}$}
\author{D. Rebuzzi$^{4}$}
\author{M. Spira$^{5}$}

\affiliation{\vspace*{0.4cm}}
\affiliation{$^{1}$ Institut f\"ur Theoretische Physik und Astrophysik, Universit\"at W\"urzburg, Emil-Hilb-Weg 22, D-97074 W\"urzburg, Germany}
\affiliation{$^{2}$ Instituto de F\'isica de Cantabria (IFCA), CSIC-Universidad de Cantabria, Santander, Spain} 
\affiliation{$^{3}$ University of Split, FESB, R. Boskovica bb, 21 000 Split, Croatia} 
\affiliation{$^{4}$ Universit\`a di Pavia and INFN Sezione di Pavia, Via A. Bassi, 6 27100 Pavia, Italy}
\affiliation{$^{5}$ Paul Scherrer Institut, CH-5232 Villigen PSI, Switzerland} 

\vspace*{1cm}     
\date{\today}

\begin{abstract}
\vspace{0.5cm}

We present an update of the branching ratios for Higgs-boson decays in
the Standard Model. We list results for all relevant branching ratios
together with corresponding uncertainties resulting from input
parameters and missing higher-order corrections. As sources of
parametric uncertainties we include the masses of the charm, bottom,
and top quarks as well as the QCD coupling constant. We compare our
results with other predictions in the literature.
%In particular, the decays of the Higgs particles to
%quark and gluon jets as well as gauge bosons are analyzed and the spread
%in the theoretical predictions due to the theoretical uncertainties and
%parametric uncertainties of the quark masses and the QCD coupling is
%discussed.
\vspace*{1.0cm}
\end{abstract}

\maketitle

\input{introduction}

\input{programs}

\input{inputparam}

\input{procedure}

\input{results}

\input{comparison}

\input{conclusions}

\section{Acknowledgments}
We thank 
J.~Baglio, 
S.~Dittmaier, 
A.~Djouadi, 
A.~Signer,
and
R.~Tanaka
for helpful discussions.

\clearpage
\appendix

\input{appendix}

\end{document}

%% file: introduction.tex
\newpage
\section{Introduction}
\label{sec:Intro}

%Motivations for this paper.
%Work done within the LHC Higgs Cross Section WG..
%etc.

One of the main goals of the LHC is the identification of the mechanism
of electroweak symmetry breaking. The most frequently investigated model
is the Higgs mechanism within the Standard Model
(SM)~\cite{higgs1,higgs2,higgs3,higgs4,higgs5} with the mass of the
Higgs boson, $\MH$, being the only unknown parameter of the SM 
relevant for the Higgs sector.
%to describe the Higgs sector in addition to the known parameters.  
All theoretical and experimental results can be expressed as a
function of $\MH$. The LEP experiments searched for the SM Higgs
boson, excluding Higgs-boson masses below $114.4\UGeV$ \cite{lephiggs}
at the 95\%~C.L., while the Tevatron established an exclusion region
of 
%$158\UGeV \le \MH \le 173 \UGeV$~\cite{Aaltonen:2011gs}. 
$156\UGeV \le \MH \le 177 \UGeV$~\cite{fermilab:2011}. 
Recently also first results on the Higgs searches at the LHC have been
published by ATLAS~\cite{Aad:2011qi} and CMS~\cite{Chatrchyan:2011tz}.%
\footnote{More recent results can be found in \citere{EPS11}.}
For a correct interpretation of the experimental data precise
calculations of the various production cross sections and the relevant
decay widths and branching ratios are necessary, including an accurate
estimate of the respective uncertainties.  To coordinate these
calculations for the Higgs boson searches at the LHC the {\em LHC
  Higgs Cross Section Working Group} was
established~\cite{LHCHiggsWG}. First results on the cross section and
branching ratio (BR) calculations for the SM Higgs boson (as well as
for Higgs bosons of the Minimal Supersymmetric Standard Model) were
recently published~\cite{LHCHiggsCrossSectionWorkingGroup:2011ti}.

In this paper we present an update of the BR calculation as well as
results for the uncertainties of the decay widths and BRs for a SM
Higgs boson.  Neglecting these uncertainties would yield in the case
of negative search results too large excluded regions of the parameter
space. In case of a Higgs boson signal these uncertainties are crucial
to perform a reliable and accurate determination of $\MH$ and the
Higgs-boson couplings~\cite{Aad:2009wy,Ball:2007zza,Duhrssen:2004cv}.
(At a future linear $\Pe^+\Pe^-$ collider the Higgs couplings can be
measured accurately in a model-independent way
\cite{AguilarSaavedra:2001rg,Djouadi:2007ik} so that the quark masses
involved in the Yukawa couplings will be extracted with much higher
precision than using QCD sum rules within the $J/\psi$ and $\Upsilon$
systems.) The uncertainties arise from two sources, the missing
higher-order corrections yield the ``theoretical'' uncertainties,
while the experimental errors on the SM input parameters, such as the
quark masses or the strong coupling constant, give rise to the
``parametric'' uncertainties.  Both types of uncertainties have to be
taken into account and combined for a reliable estimate.  We
investigate all relevant channels for the SM Higgs boson, $\Htt$,
$\Hbb$, $\Hcc$, $\Htautau$, $\Hmumu$, $\Hgg$, $\Hgaga$, $\HZga$,
$\HWW$ and $\HZZ$ (including detailed results also for the various
four-fermion final states). We present results for the total width,
$\Gatot$, as well as for various BRs.

The paper is organized as follows. In \refS{sec:Programs} we
briefly review the codes and calculations employed in our analysis. The
input parameters are summarized in
\refS{sec:Param}.  We describe in detail how the various
uncertainties are obtained in \refS{sec:Procedure}, where we also
give details about their combination to obtain the overall uncertainties.
In \refS{sec:Results} we present our main results, the uncertainty
estimates for the various decay channels (where detailed tables can be
found at the end of the paper). A brief discussion of the
differences to earlier analyses of the uncertainties is given in
\refS{sec:Comparison}. Our conclusions can be found in
\refS{sec:Conclusion}.

%%% Local Variables: 
%%% mode: latex
%%% TeX-master: "LHCHiggsBRNote"
%%% End: 

%% file: programs.tex
\section{Programs and Strategy for Branching Ratio Calculations}
\label{sec:Programs}

The branching ratios of the Higgs boson in the SM have been
determined using the programs {\HDECAY}
\cite{Djouadi:1997yw,Spira:1997dg,hdecay2} and {\Prophecy}
\cite{Bredenstein:2006rh,Bredenstein:2006ha,Prophecy4f}. In a first
step, all partial widths have been calculated as accurately as possible.
Then the branching ratios have been derived from this full set of
partial widths. Since the widths are calculated for on-shell Higgs
bosons, the results have to be used with care for a heavy Higgs boson
($\MH\gsim500\UGeV$).

\begin{itemize}
\item {\HDECAY} calculates the decay widths and branching ratios
of the Higgs boson(s) in the SM and the MSSM. For the SM it includes
all kinematically allowed channels and all relevant higher-order
QCD corrections to decays into quark pairs and into gluons. More
details are given below.

\item {\Prophecy} is a Monte Carlo event generator for $\PH \to
  \PW\PW/\PZ\PZ \to 4\Pf$ (leptonic, semi-leptonic, and hadronic) final
  states. It provides the leading-order (LO) and next-to-leading-order
  (NLO) partial widths for any possible 4-fermion final state. It
  includes the complete NLO QCD and electroweak corrections and all
  interferences at LO and NLO. In other words, it takes into account
  both the corrections to the decays into intermediate $\PW\PW$ and
  $\PZ\PZ$ states as well as their interference for final states that
  allow for both. The dominant two-loop contributions in the
  heavy-Higgs-mass limit proportional to $G_\mu^2 \MH^4$ are included
  according to Refs.~\cite{Ghinculov:1995bz,Frink:1996sv}.  Since the
  calculation is consistently performed with off-shell gauge bosons
  without any on-shell approximation, it is valid above, near, and
  below the gauge-boson pair thresholds. Like all other light quarks
  and leptons, bottom quarks are treated as massless.  Using the
  LO/NLO gauge-boson widths in the LO/NLO calculation ensures that the
  effective branching ratios of the $\PW$ and $\PZ$ bosons obtained by
  summing over all decay channels add up to one.

\item Electroweak next-to-leading order (NLO) corrections to the decays
$\PH\to\PGg\PGg$ and $\PH\to \Pg\Pg$ have been calculated in
\Brefs{Aglietti:2004ki,Aglietti:2004nj,Aglietti:2006ne,Degrassi:2004mx,
Degrassi:2005mc,Actis:2008ug,Actis:2008ts}.  They are implemented in
{\HDECAY} in form of grids based on the calculations of
\Bref{Actis:2008ug,Actis:2008ts}.
\end{itemize}

The results presented below have been obtained as follows. The Higgs total
width resulting from {\HDECAY} has been modified according to the
prescription
\begin{equation}
\Gamma_{\PH} = \Gamma^{\mathrm{HD}} - \Gamma^{\mathrm{HD}}_{\PZ\PZ} 
            - \Gamma^{\mathrm{HD}}_{\PW\PW} + \Gamma^{\mathrm{Proph.}}_{4\Pf}~,
\end{equation}
where $\Gamma_{\PH}$ is the total Higgs width, $\Gamma^{\mathrm{HD}}$
the Higgs width obtained from {\HDECAY},
$\Gamma^{\mathrm{HD}}_{\PZ\PZ}$ and $\Gamma^{\mathrm{HD}}_{\PW\PW}$
stand for the partial widths to $\PZ\PZ$ and $\PW\PW$ calculated with
{\HDECAY}, while $\Gamma^{\mathrm{Proph.}}_{4\Pf}$ represents the
partial width of $\PH\to 4\Pf$ calculated with {\Prophecy}.  The
latter can be split into the decays into $\PZ\PZ$, $\PW\PW$, and the
interference,
\begin{equation}
\Gamma^{\mathrm{Proph.}}_{4\Pf}=\Gamma_{{\PH}\to \PW^*\PW^*\to 4\Pf}
+ \Gamma_{{\PH}\to \PZ^*\PZ^*\to 4\Pf}
+ \Gamma_{\mathrm{\PW\PW/\PZ\PZ-int.}}\,.
\end{equation}

%\subsection{HDECAY}

%\subsection{Prophecy4f}

%\subsection{EW NLO Corrections}

%% file: inputparam.tex
\section{The SM Input-Parameter Set}
\label{sec:Param}

The production cross sections and decay branching ratios of the Higgs
bosons depend on a large number of SM parameters.  For the
following calculations, the input parameter set has been defined within
the {\em LHC Higgs Cross Section Working Group}%
\footnote{These parameters can be found at {\tt
    https://twiki.cern.ch/twiki/bin/view/LHCPhysics/SMInputParameter.}}
and the chosen values are listed in \refT{tab:SMinput}.
%--
\begin{table}[hb]
\renewcommand{\arraystretch}{1.4}
        \begin{center}
        \begin{tabular}{lc|lc|lc}
                \hline
%                Parameter & Value$\pm$Error \\ \hline\hline
                e mass & 0.510998910$\UMeV$ &$\mu$ mass &  105.658367$\UMeV$  &$\tau$ mass &  1776.84$\UMeV$  \\
                $\MSbar$ mass $\Mc(\Mc)$  & 1.28$\UGeV$ & $\MSbar$ mass
$\Mb(\Mb)$  & 4.16$\UGeV$  & $\MSbar$ mass $\Ms(2\UGeV)$  & 100$\UMeV$ \\
                1-loop pole mass $\Mc$  & 1.42$\UGeV$& 1-loop pole mass
$\Mb$  & 4.49$\UGeV$ & pole mass $\Mt$ & 172.5$\UGeV$ \\
                $\PW$ mass & 80.36951$\UGeV$ & NLO $\Gamma_W$ &
2.08856$\UGeV$ & &  \\
                $\PZ$ mass &91.15349$\UGeV$  & NLO $\Gamma_Z$ &
2.49581$\UGeV$  & & \\
                $\GF$ & $1.16637 \times 10^{-5}\UGeV^{-2}$ & 
$\alpha(0)$ & 1/137.0359997 &
$\alphas(\MZ^2)$ & $0.119$  \\ \hline 
        \end{tabular}
        \end{center}
        \caption{The SM input parameters 
%for particle masses and widths 
used for the branching-ratio calculations presented in this
work.}
        \label{tab:SMinput}
\end{table}

The gauge-boson masses given in \refT{tab:SMinput} are the pole masses
derived from the PDG values $\MZ=91.1876\UGeV$,
$\Gamma_\PZ=2.4952\UGeV$, $\MW=80.398\UGeV$, $\Gamma_\PW=2.141\UGeV$ and
the gauge-boson widths in \refT{tab:SMinput} are calculated at NLO from
the other input parameters.

It should be noted that for our numerical analysis we have used the
one-loop pole masses for the charm and bottom quarks and their
uncertainties, since these values do not exhibit a significant
dependence on the value of the strong coupling constant $\alphas$ in
contrast to the $\MSbar$ masses \cite{narison}.

%% file: procedure.tex
\section{Procedure for determining Uncertainties}
\label{sec:Procedure}

We included two types of uncertainties, parametric uncertainties (PU),
which originate from uncertainties in input parameters, and theoretical
uncertainties (TU), which arise from unknown contributions to the
theoretical predictions, typically missing higher orders.
Here we describe the way these uncertainties have been determined.

\subsection{Parametric uncertainties}

In order to determine the parametric uncertainties of the Higgs-decay
branching ratios we took into account the uncertainties of the input
parameters $\alphas$, $\Mc$, $\Mb$, and $\Mt$. The considered
variation of these input parameters is given in \refT{tab:inputpu}.
\begin{table}
\renewcommand{\arraystretch}{1.4}
\setlength{\arraycolsep}{1.5ex}
$$\begin{array}{cccc}
\hline
\text{\bf Parameter} & \text{\bf Central Value} & \text{\bf Uncertainty}
& \text{\bf $\MSbar$ masses $m_\Pq(m_\Pq)$} \\
\hline
  \alphas (M_Z)& 0.119 & \pm0.002\\
\Mc & 1.42\UGeV & \pm0.03\UGeV & 1.28\UGeV \\
\Mb & 4.49\UGeV & \pm0.06\UGeV & 4.16\UGeV \\
\Mt & 172.5\UGeV & \pm2.5\UGeV & 165.4\UGeV \\
\hline
\end{array}$$
\caption{Input parameters and their relative uncertainties, as used for the
uncertainty estimation of the branching ratios. The masses of the
central values correspond to the 1-loop pole masses, while the last
column contains the corresponding $\MSbar$ mass values.}
\label{tab:inputpu}
\end{table}
The variation in $\alphas$ corresponds to three times the error given in
\citeres{Bethke:2009jm,Nakamura:2010zzi}. The uncertainties for $\Mb$
and $\Mc$ are a compromise between the errors of
\citere{Nakamura:2010zzi} and the errors from the most precise
evaluations \cite{Kuhn:2007vp,Chetyrkin:2010ic,signer}.  For $\Mc$ our
error corresponds roughly to the one obtained in
\citere{Dehnadi:2011gc}.  Finally, the assumed error for $\Mt$ is about
twice the error from the most recent combination of CDF and D\O{}
\cite{:1900yx}.

We did not consider parametric uncertainties resulting from experimental
errors on $\GF$, $\MZ$, $\MW$ and the lepton masses, because their
impact is below one per mille. We also did not include uncertainties for
the light quarks u, d, s as the corresponding branching ratios are very
small and the impact on other branching ratios is negligible. Since we
used $\GF$ to fix the electromagnetic coupling $\alpha(\MZ)$,
uncertainties in the hadronic vacuum polarisation do not matter.

Given the uncertainties in the parameters, the parametric uncertainties
have been determined as follows. For each parameter
$p=\alphas,\Mc,\Mb,\Mt$ we have calculated the Higgs branching ratios
for $p$, $p+\Delta p$ and $p-\Delta p$, while all other parameters have
been left at their central values. The error on each branching ratio
has then been determined by
\begin{eqnarray}
\Delta^p_+ \br &=&  \max \{\br(p+\Delta p),\br(p),\br(p-\Delta p)\} -
\br(p)\nonumber\\
\Delta^p_- \br &=&  \br(p) -\min \{\br(p+\Delta p),\br(p),\br(p-\Delta p)\}.
\end{eqnarray}
The total parametric errors have been obtained by adding the parametric
errors from the four parameter variations in quadrature. This procedure
ensures that the branching ratios add up to unity for all parameter
variations individually.

The uncertainties of the partial and total decay widths have been
obtained in an analogous way,
\begin{eqnarray}
\Delta^p_+ \Gamma &=&  \max \{\Gamma(p+\Delta
p),\Gamma(p),\Gamma(p-\Delta p)\} - \Gamma(p)\nonumber\\
\Delta^p_- \Gamma &=&  \Gamma(p) -\min \{\Gamma(p+\Delta
p),\Gamma(p),\Gamma(p-\Delta p)\}.
\end{eqnarray}
where $\Gamma$ denotes the partial decay width for each considered decay
channel or the total width, respectively.  The total parametric errors
have been derived by adding the individual parametric errors in
quadrature.

\subsection{Theoretical uncertainties}

The second type of uncertainties for the Higgs branching ratios
results from approximations in the theoretical calculations, the
dominant effects being due to missing higher orders. Since the decay
widths have been calculated with \HDECAY\ and \Prophecy\ the missing
contributions in these codes are relevant. For QCD corrections the
uncertainties have been estimated by the scale dependence of the
widths resulting from a variation of the scale up and down by a factor
2 or from the size of known omitted corrections. For electroweak
corrections the missing higher orders have been estimated based on the
known structure and size of the NLO corrections. For cases where
\HDECAY\ takes into account the known NLO corrections only
approximatively the accuracy of these approximations has been used.
The estimated relative theoretical uncertainties for the partial
widths resulting from missing higher-order corrections are summarized
in \refT{tab:uncertainty}. The corresponding uncertainty for the total
width is obtained by adding the uncertainties for the partial widths
linearly.
\begin{table}[ht]
\begin{center}
\renewcommand{\arraystretch}{1.4}
\setlength{\tabcolsep}{1.5ex}
\begin{tabular}{lllll}
\hline
\textbf{Partial Width} & \textbf{QCD} & \textbf{Electroweak} & \textbf{Total} \\
\hline
 $\PH \to \PQb\bar\PQb/\PQc\bar\PQc$ &    $\sim 0.1\%$
&     $\sim 1$--$2\%$ for $\MH \lsim 135\UGeV$     &      $\sim 2 \%$\\
%\hline
$\PH\to \tau^+ \tau^-/\mu^+\mu^-$ & & $\sim1$--$2\%$  for $\MH \lsim 135\UGeV$ &       $\sim 2 \%$ \\
%\hline
$\PH \to \PQt\bar\PQt$ & $\lsim 5\%$& 
      ${\lsim  2}$--${5\%}$   for $\MH < 500 \UGeV$     &       $\sim5\%$ \\
& &      $\sim 0.1 (\frac{\MH}{1\UTeV})^4$ for $\MH > 500\UGeV$  &       $\sim5$--$10\%$ \\      
%\hline
$\PH \to \Pg\Pg$ & ${\sim 3\%}$   & 
$\sim 1\%$   &   $\sim3\%$\\
%\hline
$\PH \to \PGg \PGg$  & ${<1\%}$ & $<1\%$    &  $\sim1\%$ \\
%\hline
$\PH \to \PZ \PGg$  & ${<1\%}$ & $\sim 5\%$    &  $\sim5\%$ \\
$\PH \to \PW\PW/\PZ\PZ\to4\Pf$ & $<0.5\%$ &   $\sim 0.5\%$ for $\MH < 500\UGeV$ &  $\sim0.5\%$\\
$\phantom{\PH\to    \to 4\Pf}$             && $\sim 0.17 (\frac{\MH}{1\UTeV})^4$ for $\MH > 500\UGeV$
&        $\sim0.5$--$15\%$                          \\ 
\hline
\end{tabular}
\end{center}
\caption{Estimated theoretical uncertainties from missing higher orders.}
\label{tab:uncertainty}
\end{table}

Specifically, the uncertainties of the partial widths calculated with
\HDECAY\ are obtained as follows: For the decays $\PH \to \PQb\bar\PQb,
\PQc\bar\PQc$, \HDECAY\ includes the complete massless QCD corrections
up to and including NNNNLO, with a corresponding scale dependence of
about $0.1\%$
\cite{Gorishnii:1990zu,Gorishnii:1991zr,Kataev:1993be,Surguladze:1994gc,
Larin:1995sq,Chetyrkin:1995pd,Chetyrkin:1996sr,Baikov:2005rw}. The NLO
electroweak corrections
\cite{Fleischer:1980ub,Bardin:1990zj,Dabelstein:1991ky,Kniehl:1991ze}
are included in the approximation for small Higgs masses
\cite{Djouadi:1991uf} which has an accuracy of about $1{-}2\%$ for $\MH
< 135\UGeV$. The same applies to the electroweak corrections to $\PH \to
\tau^+ \tau^-$.  For Higgs decays into top quarks \HDECAY\ includes the
complete NLO QCD corrections
\cite{Braaten:1980yq,Sakai:1980fa,Inami:1980qp,Gorishnii:1983cu,
Drees:1989du,Drees:1990dq,Drees:1991dq} interpolated to the
large-Higgs-mass results at NNNNLO far above the threshold
\cite{Gorishnii:1990zu,Gorishnii:1991zr,Kataev:1993be,Surguladze:1994gc,
Larin:1995sq,Chetyrkin:1995pd,Chetyrkin:1996sr,Baikov:2005rw}.  The
corresponding scale dependence is below $5\%$.  Only the NLO electroweak
corrections due to the self-interaction of the Higgs boson are included,
and the neglected electroweak corrections amount to about $2{-}5\%$ for
$\MH < 500\UGeV$, where $5\%$ refers to the region near the
$\PQt\bar\PQt$ threshold and $2\%$ to Higgs masses far above.  For $\MH
> 500\UGeV$ higher-order heavy-Higgs corrections
\cite{Ghinculov:1994se,Ghinculov:1995err,Durand:1994pk,Durand:1994err,
Durand:1994pw,Borodulin:1996br} dominate the error, resulting in an
uncertainty of about $0.1\times(\MH/1\UTeV)^4$ for $\MH > 500\UGeV$.
For $\PH \to \Pg\Pg$, \HDECAY\ uses the NLO
\cite{Inami:1982xt,Djouadi:1991tka,Spira:1995rr}, NNLO
\cite{Chetyrkin:1997iv}, and NNNLO  \cite{Baikov:2006ch} QCD corrections
in the limit of heavy top quarks.  The uncertainty from the scale
dependence at NNNLO is about $3\%$.  The NLO electroweak corrections are
included via an interpolation based on a grid from \Bref{Actis:2008ts};
the uncertainty from missing higher-order electroweak corrections is
estimated to be $1\%$.  For the decay $\PH\to\PGg \PGg$, \HDECAY\
includes the full NLO QCD corrections
\cite{Zheng:1990qa,Djouadi:1990aj,Dawson:1992cy,Djouadi:1993ji,
Melnikov:1993tj,Inoue:1994jq,Spira:1995rr} and a grid from
\Bref{Actis:2008ug,Actis:2008ts} for the NLO electroweak corrections.
Missing higher orders are estimated to be below $1\%$.  The contribution
of the $\PH \rightarrow \PGg \Pe^{+}\Pe^{-}$ decay via virtual photon
conversion, evaluated in \Bref{Firan:2007tp} is not taken into account
in the following results. Its correct treatment and its inclusion in
HDECAY are in progress.\footnote{The contribution of $\PH \rightarrow
\PGg \Pe^{+}\Pe^{-}$ is part of the QED corrections to $\PH \rightarrow
\PGg \PGg$ which are expected to be small in total.} The partial decay
width $H\to \PZ \PGg$ is included in \HDECAY at LO including the virtual
$W$, top, bottom and $\tau$ loop contributions. The QCD corrections are
small in the intermediate Higgs mass range \cite{Spira:1991tj} and can
thus safely be neglected. The associated theoretical uncertainty ranges
at the level below one per cent. The electroweak corrections to this
decay mode are unknown and thus imply a theoretical uncertainty of about
5\% in the intermediate Higgs mass range.

The decays $\PH \to \PW\PW/\PZ\PZ\to4\Pf$ are based on \Prophecy, which
includes the complete NLO QCD and electroweak corrections with all
interferences and leading two-loop heavy-Higgs corrections.  For small
Higgs-boson masses the missing higher-order corrections are estimated to
roughly $0.5\%$. For $\MH > 500\UGeV$ higher-order heavy-Higgs
corrections dominate the error leading to an uncertainty of about
$0.17\times(\MH/1\UTeV)^4$.

Based on the error estimates for the partial widths in
\refT{tab:uncertainty}, the theoretical uncertainties for the
branching ratios are determined as follows. For the partial widths
$\PH \to \PQb\bar\PQb,\PQc\bar\PQc,\tau^+ \tau^-, \Pg\Pg, \PGg \PGg$
the total uncertainty given in \refT{tab:uncertainty} is used.  For
$\PH \to \PQt\bar\PQt$ and $\PH\to \PW\PW/\PZ\PZ\to4\Pf$, the total
uncertainty is used for $\MH<500\UGeV$, while for higher Higgs masses
the QCD and electroweak uncertainties are added linearly.  Then the
shifts of all branching ratios are calculated resulting from the
scaling of an individual partial width by the corresponding relative
error (since each branching ratio depends on all partial widths,
scaling a single partial width modifies all branching ratios). This is
done by scaling each partial width separately while fixing all others
to their central values, resulting in individual theoretical
uncertainties of each branching ratio.  However, since the errors for
all $\PH\to \PW\PW/\PZ\PZ\to4\Pf$ decays are correlated for
$\MH>500\UGeV$ or small below, we only consider the simultaneous
scaling of all 4-fermion partial widths. The thus obtained individual
theoretical uncertainties for the branching ratios are combined
linearly to obtain the total theoretical uncertainties.

\vspace{1ex}

Finally, the total uncertainties are obtained by adding linearly the
total parametric uncertainties and the total theoretical uncertainties.

%%% Local Variables: 
%%% mode: latex
%%% TeX-master: ../LHCHiggsBRNote.tex
%%% TeX-master: t
%%% End: 

%% file: results.tex
\section{Results}
\label{sec:Results}

\begin{figure}%[htb!]
\centerline{\includegraphics[width=14.7cm]{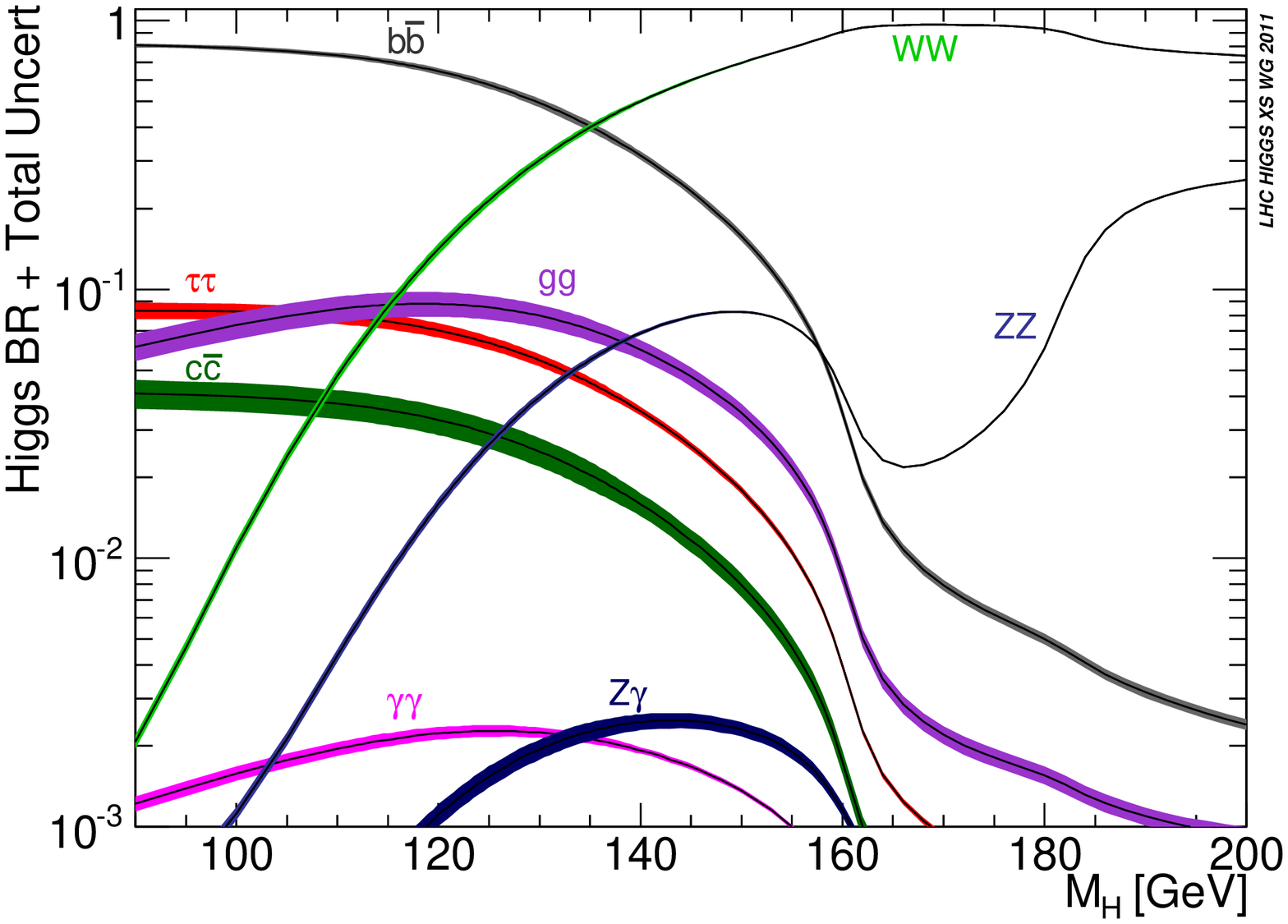}}
\vspace*{-0.3cm}
\caption{Higgs branching ratios and their uncertainties for the low mass range.}
\label{fig:BRTotUnc_lm}
\end{figure}

\begin{figure}%[htb!]
\centerline{\includegraphics[width=14.7cm]{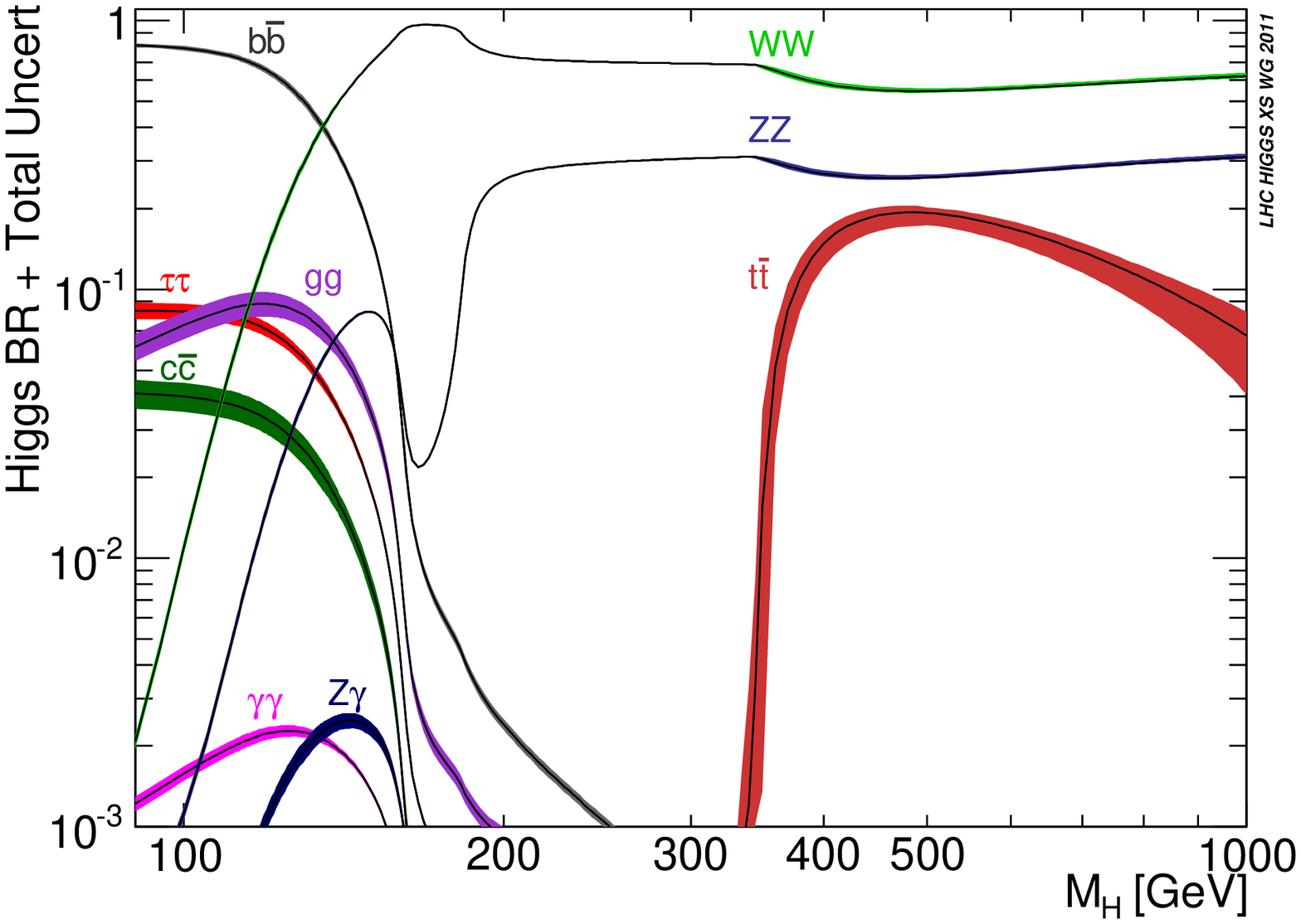}}
\vspace*{-0.3cm}
\caption{Higgs branching ratios and their uncertainties for the full mass range.}
\label{fig:BRTotUnc}
\end{figure}

\begin{figure}%[htb!]
\centerline{\includegraphics[width=14.3cm]{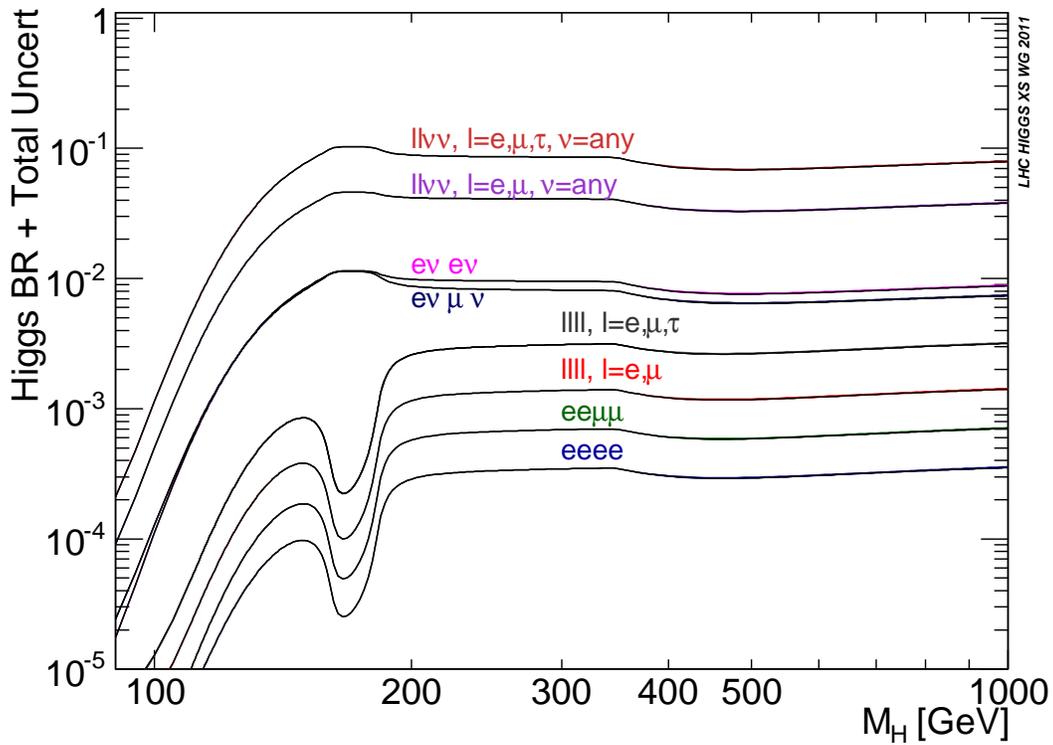}}
\vspace*{-0.3cm}
\caption{Higgs branching ratios for the different $\PH \to 4\Pl$ and
$\PH \to 2\Pl2\nu$ final states and their uncertainties for the full
  mass range.} 
\label{fig:BRTotUnc_VVlept}
\end{figure}

\begin{figure}%[htb!]
\centerline{\includegraphics[width=14.3cm]{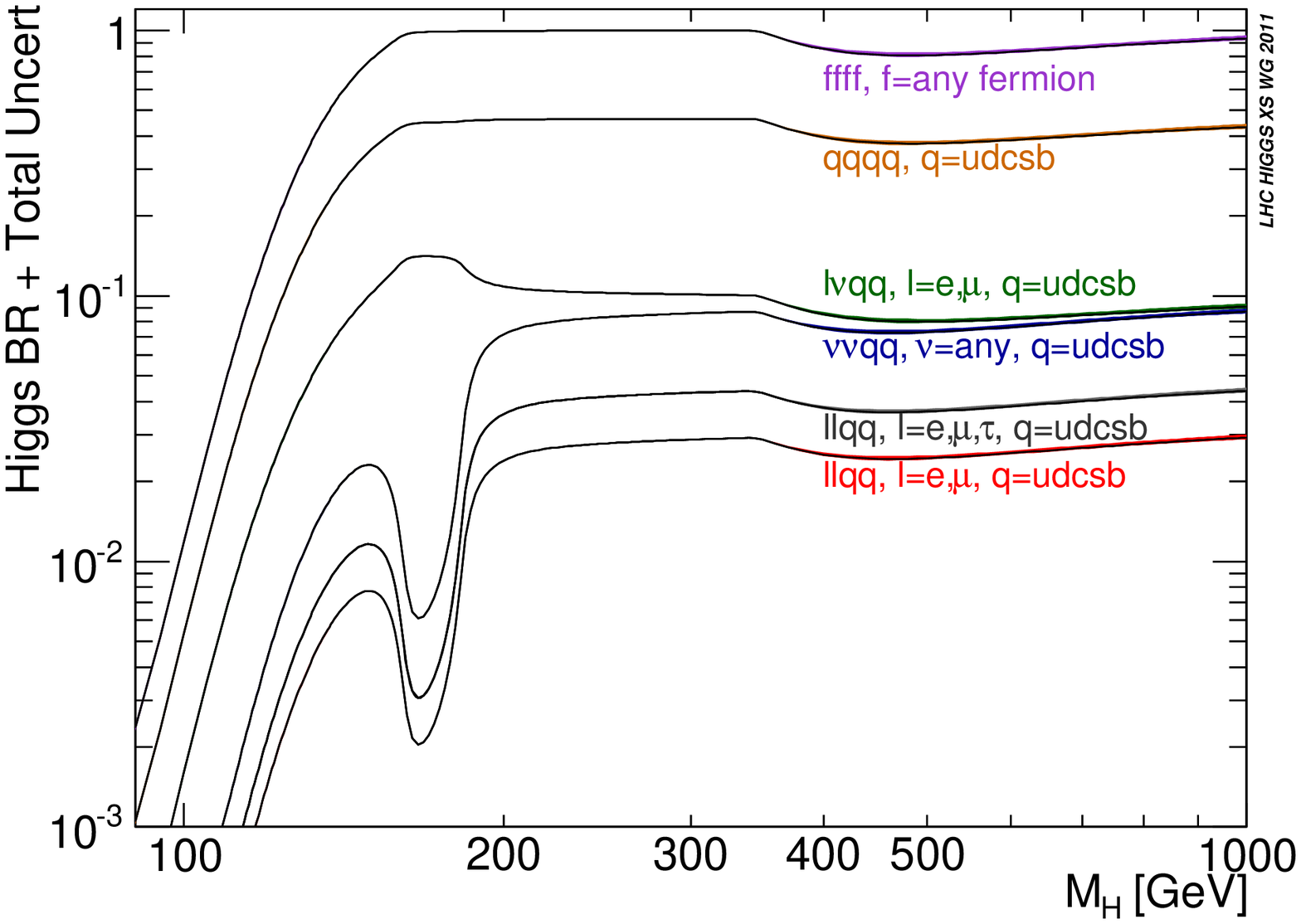}}
\vspace*{-0.3cm}
\caption{Higgs branching ratios for $\PH \to 4\Pq$, $\PH \to 4\Pf$ and
$\PH \to 2\Pq2\Pl, 2\Pq \Pl \nu, 2\Pq 2 \nu$ final states and their
  uncertainties for the full mass range.} 
\label{fig:BRTotUnc_VVhad}
\end{figure}

In this Section the results of the SM Higgs branching ratios, calculated
according to the procedure described above, are shown and discussed.
\refF{fig:BRTotUnc_lm} shows the SM Higgs branching ratios in the low
mass range, $100 \UGeV \le \MH \le 200 \UGeV$ as solid lines. The (coloured)
bands around the lines show the respective uncertainties, estimated
considering both the theoretical and the 
parametric uncertainty sources (as discussed in \refS{sec:Procedure}).
The same results, but now for the ``full'' mass range,
$100 \UGeV \le \MH \le 1000 \UGeV$, are shown in \refF{fig:BRTotUnc}. 
More detailed results on the decays $\HWW$ and $\HZZ$ with the subsequent
decay to $4\Pf$ are presented in \refFs{fig:BRTotUnc_VVlept}
and~\ref{fig:BRTotUnc_VVhad}. 
The largest ``visible'' uncertainties can are found for the channels
$\Htautau$, $\Hgg$, $\Hcc$ and $\Htt$, see below.

\refTs{BR-2fa}--\ref{BR-2gbe}, which can be found at the end of the paper,
show the branching ratios for the Higgs two-body fermionic and bosonic 
final states, together with their total uncertainties, estimated
as discussed in \refS{sec:Procedure}.%
\footnote{The value $0.0\%$ means that the uncertainty is below
  $0.05\%$.}
represents  \refTs{BR-2gba}--\ref{BR-2gbe}
also contain the total Higgs width $\Gamma_{\PH}$ in the last column.
More detailed results for four representative Higgs-boson masses are
given in \refT{BR1}. Here we show the BR, the PU separately for the
four parameters as given in \refT{tab:inputpu}, the total PU, the
theoretical uncertainty TU as well as the total uncertainty on the
Higgs branching ratios. 
The TU are most relevant for the $\Hgg$, $\HZga$ and $\Htt$ branching
ratios, reaching $\cal O$(10\%). For the $\Hbb$, $\Hcc$ and $\Htautau$
branching ratios they remain below a few percent. 
PU are relevant mostly for the $\Hcc$ and $\Hgg$ branching ratios,
reaching up to
$\cal O$(10\%) and $\cal O$(5\%), respectively. 
They are mainly induced by the parametric uncertainties in $\alphas$
and $\Mc$. 
The PU resulting from $\Mb$ affect the $\br(\Hbb)$ at the level of $3\%$,
and the PU from $\Mt$ influences in particular the $\br(\Htt)$ near the
$\PQt\bar{\PQt}$ threshold.
For the $\Hgaga$ channel the total uncertainty can reach up to about
$5\%$ in the relevant mass range.
Both TU and PU on the important channels $\HZZ$ and $\HWW$ 
remain at the level of 1\% over the full mass range, giving rise to a total
uncertainty below 3\% for $\MH > 135 \UGeV$.

%%%%%%%%%%%%%%%%%%%%%%%%%% T A B L E %%%%%%%%%%%%%%%%%%%%%%%%%%%%%%%%%%%%%%%%
\begin{table}[htb]
\renewcommand{\arraystretch}{1.5}
\setlength{\tabcolsep}{1.5ex}
\begin{tabular}{lccrrrrrrr}
\hline
Channel & $\MH [\UGeV]$  &  BR & $\Delta \Mc$ & $\Delta \Mb$ & $\Delta \Mt$ & $\Delta \alphas$ & PU  & TU & Total \\
\hline
& 120  & 6.48E-01&$^{-0.2\%}_{+0.2\%}$  & $^{+1.1\%}_{-1.2\%}$  & $^{+0.0\%}_{-0.0\%}$  & $^{-1.0\%}_{+0.9\%}$  & $^{+1.5\%}_{-1.5\%}$  &$^{+1.3\%}_{-1.3\%}$   & $^{+2.8\%}_{-2.8\%}$   \\ 
& 150  & 1.57E-01&$^{-0.1\%}_{+0.1\%}$  & $^{+2.7\%}_{-2.7\%}$  & $^{+0.1\%}_{-0.1\%}$  & $^{-2.2\%}_{+2.1\%}$  & $^{+3.4\%}_{-3.5\%}$  &$^{+0.6\%}_{-0.6\%}$   & $^{+4.0\%}_{-4.0\%}$   \\[-1ex]
\raisebox{1.5ex}{$\Hbb$}
& 200  & 2.40E-03&$^{-0.0\%}_{+0.0\%}$  & $^{+3.2\%}_{-3.2\%}$  & $^{+0.0\%}_{-0.1\%}$  & $^{-2.5\%}_{+2.5\%}$  & $^{+4.1\%}_{-4.1\%}$  &$^{+0.5\%}_{-0.5\%}$   & $^{+4.6\%}_{-4.6\%}$   \\ 
& 500  & 1.09E-04&$^{-0.0\%}_{+0.0\%}$  & $^{+3.2\%}_{-3.2\%}$  & $^{+0.1\%}_{-0.1\%}$  & $^{-2.8\%}_{+2.8\%}$  & $^{+4.3\%}_{-4.3\%}$  &$^{+3.0\%}_{-1.1\%}$   & $^{+7.2\%}_{-5.4\%}$   \\ 
\hline

& 120  & 7.04E-02        &$^{-0.2\%}_{+0.2\%}$  & $^{-2.0\%}_{+2.1\%}$  & $^{+0.1\%}_{-0.1\%}$  & $^{+1.4\%}_{-1.3\%}$  & $^{+2.5\%}_{-2.4\%}$ &$^{+3.6\%}_{-3.6\%}$  & $^{+6.1\%}_{-6.0\%}$       \\ 
& 150  & 1.79E-02        &$^{-0.1\%}_{+0.1\%}$  & $^{-0.5\%}_{+0.5\%}$  & $^{+0.1\%}_{-0.1\%}$  & $^{+0.3\%}_{-0.3\%}$  & $^{+0.6\%}_{-0.6\%}$ &$^{+2.5\%}_{-2.5\%}$  & $^{+3.0\%}_{-3.1\%}$       \\[-1ex] 
\raisebox{1.5ex}{$\Htautau$} 
& 200  & 2.87E-04        &$^{-0.0\%}_{+0.0\%}$  & $^{-0.0\%}_{+0.0\%}$  & $^{+0.0\%}_{-0.1\%}$  & $^{+0.0\%}_{-0.0\%}$  & $^{+0.0\%}_{-0.1\%}$ &$^{+2.5\%}_{-2.5\%}$  & $^{+2.5\%}_{-2.6\%}$       \\ 
& 500  & 1.53E-05        &$^{-0.0\%}_{+0.0\%}$  & $^{-0.0\%}_{+0.0\%}$  & $^{+0.1\%}_{-0.1\%}$  & $^{-0.1\%}_{+0.0\%}$  & $^{+0.1\%}_{-0.1\%}$ &$^{+5.0\%}_{-3.1\%}$  & $^{+5.0\%}_{-3.2\%}$       \\ 
\hline

& 120  & 2.44E-04        &$^{-0.2\%}_{+0.2\%}$  & $^{-2.0\%}_{+2.1\%}$  & $^{+0.1\%}_{-0.1\%}$  & $^{+1.4\%}_{-1.3\%}$  & $^{+2.5\%}_{-2.5\%}$ &$^{+3.9\%}_{-3.9\%}$  & $^{+6.4\%}_{-6.3\%}$       \\ 
& 150  & 6.19E-05        &$^{-0.0\%}_{+0.0\%}$  & $^{-0.5\%}_{+0.5\%}$  & $^{+0.1\%}_{-0.1\%}$  & $^{+0.3\%}_{-0.3\%}$  & $^{+0.6\%}_{-0.6\%}$ &$^{+2.5\%}_{-2.5\%}$  & $^{+3.1\%}_{-3.2\%}$       \\[-1ex] 
\raisebox{1.5ex}{$\Hmumu$}
& 200  & 9.96E-07        &$^{-0.0\%}_{-0.0\%}$  & $^{-0.0\%}_{+0.0\%}$  & $^{+0.1\%}_{-0.1\%}$  & $^{+0.0\%}_{-0.0\%}$  & $^{+0.1\%}_{-0.1\%}$ &$^{+2.5\%}_{-2.5\%}$  & $^{+2.6\%}_{-2.6\%}$       \\ 
& 500  & 5.31E-08        &$^{-0.0\%}_{+0.0\%}$  & $^{-0.0\%}_{+0.0\%}$  & $^{+0.1\%}_{-0.1\%}$  & $^{-0.0\%}_{+0.0\%}$  & $^{+0.1\%}_{-0.1\%}$ &$^{+5.0\%}_{-3.1\%}$  & $^{+5.1\%}_{-3.1\%}$       \\ 
\hline

& 120  & 3.27E-02        &$^{+6.0\%}_{-5.8\%}$  & $^{-2.1\%}_{+2.2\%}$  & $^{+0.1\%}_{-0.1\%}$  & $^{-5.8\%}_{+5.6\%}$  & $^{+8.5\%}_{-8.5\%}$  &$^{+3.8\%}_{-3.7\%}$ & $^{+12.2\%}_{-12.2\%}$      \\ 
& 150  & 7.93E-03        &$^{+6.2\%}_{-6.0\%}$  & $^{-0.6\%}_{+0.6\%}$  & $^{+0.1\%}_{-0.1\%}$  & $^{-6.9\%}_{+6.8\%}$  & $^{+9.2\%}_{-9.2\%}$  &$^{+0.6\%}_{-0.6\%}$ & $^{+9.7\%}_{-9.7\%}$        \\[-1ex] 
\raisebox{1.5ex}{$\Hcc$} 
& 200  & 1.21E-04        &$^{+6.2\%}_{-6.1\%}$  & $^{-0.2\%}_{+0.1\%}$  & $^{+0.1\%}_{-0.2\%}$  & $^{-7.2\%}_{+7.2\%}$  & $^{+9.5\%}_{-9.5\%}$  &$^{+0.5\%}_{-0.5\%}$ & $^{+10.0\%}_{-10.0\%}$      \\ 
& 500  & 5.47E-06        &$^{+6.2\%}_{-6.0\%}$  & $^{-0.1\%}_{+0.1\%}$  & $^{+0.1\%}_{-0.1\%}$  & $^{-7.6\%}_{+7.6\%}$  & $^{+9.8\%}_{-9.7\%}$  &$^{+3.0\%}_{-1.1\%}$ & $^{+12.8\%}_{-10.7\%}$      \\ 
\hline

& 350  & 1.56E-02        &$^{+0.0\%}_{+0.0\%}$  & $^{-0.0\%}_{+0.0\%}$  & $^{~-78.6\%}_{+120.9\%}$  & $^{+0.9\%}_{-0.9\%}$ & $^{+120.9\%} _{~-78.6\%}$  &$^{~+6.9\%}_{-12.7\%}$  & $^{+127.8\%} _{~-91.3\%}$      \\ 
& 360  & 5.14E-02        &$^{-0.0\%}_{-0.0\%}$  & $^{-0.0\%}_{+0.0\%}$  & $^{-36.2\%}_{+35.6\%}$   & $^{+0.7\%}_{-0.7\%}$ & $^{+35.6\%}  _{-36.2\%}$  &$^{~+6.6\%}_{-12.2\%}$  & $^{+42.2\%}  _{-48.4\%}$      \\[-1ex] 
\raisebox{1.5ex}{$\Htt$}
& 400  & 1.48E-01        &$^{+0.0\%}_{+0.0\%}$  & $^{-0.0\%}_{+0.0\%}$  & $^{-6.8\%} _{+6.2\%}$    & $^{+0.4\%}_{-0.3\%}$ & $^{+6.2\%}   _{-6.8\%}$   &$^{~+5.9\%}_{-11.1\%}$  & $^{+12.2\%}  _{-17.8\%}$      \\ 
& 500  & 1.92E-01        &$^{-0.0\%}_{+0.0\%}$  & $^{-0.0\%}_{+0.0\%}$  & $^{-0.3\%} _{+0.1\%}$    & $^{+0.1\%}_{-0.2\%}$ & $^{+0.1\%}   _{-0.3\%}$   &$^{+4.5\%}_{-9.5\%}$   & $^{+4.6\%}   _{-9.8\%}$       \\ 
\hline

& 120  & 8.82E-02        &$^{-0.2\%}_{+0.2\%}$  & $^{-2.2\%}_{+2.2\%}$  & $^{-0.2\%}_{+0.2\%}$   & $^{+5.7\%}_{-5.4\%}$ & $^{+6.1\%}_{-5.8\%}$  &$^{+4.5\%}_{-4.5\%}$  & $^{+10.6\%}_{-10.3\%}$   \\ 
& 150  & 3.46E-02        &$^{-0.1\%}_{+0.1\%}$  & $^{-0.7\%}_{+0.6\%}$  & $^{-0.3\%}_{+0.3\%}$   & $^{+4.4\%}_{-4.2\%}$ & $^{+4.4\%}_{-4.3\%}$  &$^{+3.5\%}_{-3.5\%}$  & $^{+7.9\%} _{-7.8\%}$    \\[-1ex] 
\raisebox{1.5ex}{$\Hgg$}
& 200  & 9.26E-04        &$^{-0.0\%}_{-0.0\%}$  & $^{-0.1\%}_{+0.1\%}$  & $^{-0.6\%}_{+0.6\%}$   & $^{+3.9\%}_{-3.8\%}$ & $^{+3.9\%}_{-3.9\%}$  &$^{+3.7\%}_{-3.7\%}$  & $^{+7.6\%} _{-7.6\%}$    \\ 
& 500  & 6.04E-04        &$^{-0.0\%}_{+0.0\%}$  & $^{-0.0\%}_{+0.0\%}$  & $^{+1.6\%} _{-1.6\%}$  & $^{+3.4\%}_{-3.3\%}$ & $^{+3.7\%}_{-3.7\%}$  &$^{+6.2\%}_{-4.3\%}$  & $^{+9.9\%} _{-7.9\%}$    \\ 
\hline

& 120  & 2.23E-03        &$^{-0.2\%}_{+0.2\%}$  & $^{-2.0\%}_{+2.1\%}$  & $^{+0.0\%} _{+0.0\%}$  & $^{+1.4\%}    _{-1.3\%}$   & $^{+2.5\%}_{-2.4\%}$   &$^{+2.9\%}_{-2.9\%}$  & $^{+5.4\%} _{-5.3\%}$   \\ 
& 150  & 1.37E-03        &$^{+0.0\%} _{+0.1\%}$ & $^{-0.5\%}_{+0.5\%}$  & $^{+0.1\%} _{-0.0\%}$ & $^{+0.3\%}    _{-0.3\%}$   & $^{+0.6\%}_{-0.6\%}$   &$^{+1.6\%}_{-1.5\%}$  & $^{+2.1\%} _{-2.1\%}$    \\[-1ex] 
\raisebox{1.5ex}{$\Hgaga$}
& 200  & 5.51E-05        &$^{-0.0\%}_{-0.0\%}$  & $^{-0.0\%}_{+0.0\%}$  & $^{+0.1\%} _{-0.1\%}$ & $^{+0.0\%}    _{-0.0\%}$   & $^{+0.1\%}_{-0.1\%}$   &$^{+1.5\%}_{-1.5\%}$  & $^{+1.6\%} _{-1.6\%}$    \\ 
& 500  & 3.12E-07        &$^{-0.0\%}_{+0.0\%}$  & $^{-0.0\%}_{+0.0\%}$  & $^{+8.0\%} _{-6.5\%}$ & $^{-0.7\%}   _{+0.7\%}$    & $^{+8.0\%}_{-6.6\%}$   &$^{+4.0\%}_{-2.1\%}$  & $^{+11.9\%}_{~-8.7\%}$    \\ 
\hline

& 120  & 1.11E-03        &$^{-0.3\%}_{+0.2\%}$  & $^{-2.1\%}_{+2.1\%}$  & $^{+0.0\%} _{-0.1\%}$ & $^{+1.4\%}    _{-1.4\%}$   & $^{+2.5\%}_{-2.5\%}$   &$^{+6.9\%}_{-6.8\%}$  & $^{+9.4\%} _{-9.3\%}$    \\ 
& 150  & 2.31E-03        &$^{-0.1\%}_{+0.0\%}$  & $^{-0.6\%}_{+0.5\%}$  & $^{+0.0\%} _{-0.1\%}$ & $^{+0.2\%}    _{-0.3\%}$   & $^{+0.5\%}_{-0.6\%}$   &$^{+5.5\%}_{-5.5\%}$  & $^{+6.0\%} _{-6.2\%}$    \\[-1ex] 
\raisebox{1.5ex}{$\HZga$}
& 200  & 1.75E-04        &$^{-0.0\%}_{-0.0\%}$  & $^{-0.0\%}_{+0.0\%}$  & $^{+0.0\%} _{-0.1\%}$ & $^{+0.0\%}    _{-0.0\%}$   & $^{+0.0\%}_{-0.1\%}$   &$^{+5.5\%}_{-5.5\%}$  & $^{+5.5\%} _{-5.6\%}$    \\ 
& 500  & 7.58E-06        &$^{-0.0\%}_{+0.0\%}$  & $^{-0.0\%}_{+0.0\%}$  & $^{+0.8\%} _{-0.6\%}$ & $^{-0.0\%}   _{+0.0\%}$    & $^{+0.8\%}_{-0.6\%}$   &$^{+8.0\%}_{-6.1\%}$  & $^{+8.7\%} _{-6.7\%}$    \\ 
\hline

& 120  & 1.41E-01        &$^{-0.2\%}_{+0.2\%}$  & $^{-2.0\%}_{+2.1\%}$  & $^{-0.0\%}_{+0.0\%}$  & $^{+1.4\%}    _{-1.4\%}$   & $^{+2.5\%}_{-2.5\%}$   &$^{+2.2\%}_{-2.2\%}$  & $^{+4.8\%} _{-4.7\%}$    \\ 
& 150  & 6.96E-01        &$^{-0.1\%}_{+0.1\%}$  & $^{-0.5\%}_{+0.5\%}$  & $^{-0.0\%}_{+0.0\%}$  & $^{+0.3\%}    _{-0.3\%}$   & $^{+0.6\%}_{-0.6\%}$   &$^{+0.3\%}_{-0.3\%}$  & $^{+0.9\%} _{-0.8\%}$    \\[-1ex] 
\raisebox{1.5ex}{$\HWW$}
& 200  & 7.41E-01        &$^{-0.0\%}_{-0.0\%}$  & $^{-0.0\%}_{+0.0\%}$  & $^{-0.0\%}_{+0.0\%}$  & $^{+0.0\%}    _{-0.0\%}$   & $^{+0.0\%}_{-0.0\%}$   &$^{+0.0\%}_{-0.0\%}$  & $^{+0.0\%} _{-0.0\%}$    \\ 
& 500  & 5.46E-01        &$^{-0.0\%}_{+0.0\%}$  & $^{-0.0\%}_{+0.0\%}$  & $^{+0.1\%} _{-0.0\%}$ & $^{-0.0\%}   _{+0.0\%}$    & $^{+0.1\%}_{-0.1\%}$   &$^{+2.3\%}_{-1.1\%}$  & $^{+2.4\%} _{-1.1\%}$    \\ 
\hline

& 120  & 1.59E-02        &$^{-0.2\%}_{+0.2\%}$  & $^{-2.0\%}_{+2.1\%}$  & $^{-0.0\%}_{+0.0\%}$  & $^{+1.4\%}    _{-1.4\%}$   & $^{+2.5\%}_{-2.5\%}$   &$^{+2.2\%}_{-2.2\%}$  & $^{+4.8\%} _{-4.7\%}$    \\ 
& 150  & 8.25E-02        &$^{-0.1\%}_{+0.1\%}$  & $^{-0.5\%}_{+0.5\%}$  & $^{+0.0\%} _{+0.0\%}$  & $^{+0.3\%}    _{-0.3\%}$   & $^{+0.6\%}_{-0.6\%}$   &$^{+0.3\%}_{-0.3\%}$  & $^{+0.9\%} _{-0.8\%}$   \\[-1ex] 
\raisebox{1.5ex}{$\HZZ$}
& 200  & 2.55E-01        &$^{-0.0\%}_{+0.0\%}$  & $^{-0.0\%}_{+0.0\%}$  & $^{+0.0\%} _{-0.0\%}$ & $^{+0.0\%}    _{-0.0\%}$   & $^{+0.0\%}_{-0.0\%}$   &$^{+0.0\%}_{-0.0\%}$  & $^{+0.0\%} _{-0.0\%}$    \\ 
& 500  & 2.61E-01        &$^{+0.0\%}_{-0.0\%}$  & $^{-0.0\%}_{+0.0\%}$  & $^{+0.0\%} _{+0.0\%}$  & $^{-0.0\%}   _{+0.0\%}$    & $^{+0.1\%}_{-0.0\%}$   &$^{+2.3\%}_{-1.1\%}$  & $^{+2.3\%} _{-1.1\%}$   \\ 
\hline

\end{tabular}
\caption{SM Higgs branching ratios and their relative parametric (PU),
  theoretical (TU) and total uncertainties for a selection of Higgs
  masses. For PU, all the single contributions are shown. For these
  four columns, the upper percentage value (with its sign) refers 
  to the positive variation of the parameter, while the lower one
  refers to the negative variation of 
  the parameter. Results for the full mass range, including the total
  uncertainties, are listed in Tables at the  
end of the document.}
\label{BR1}
\end{table}
%%%%%%%%%%%%%%%%%%%%%%%%%% T A B L E %%%%%%%%%%%%%%%%%%%%%%%%%%%%%%%%%%%%%%%%

Finally, \refTs{BR-4f1a}--\ref{BR-4f1e} and
\refTs{BR-4f2a}--\ref{BR-4f2e}, to be found at the end of the paper,
list the branching ratios for the most relevant Higgs decays into
four-fermion final states. The right column in these Tables shows the
total relative uncertainties on these branching ratios in percentage.
These are practically equal for all the $\PH \to 4\Pf$ branching
ratios and the same as those for $\PH \to \PW\PW/\PZ\PZ$.
It should be noted that the charge-conjugate state is not included for
$\PH \to \Pl\nu \Pq\Pqb$.

We would like to remark that, when possible, the branching ratios for
Higgs into four fermions, explicitly calculated and listed in 
\refTs{BR-4f1a}--\ref{BR-4f1e}, should be preferred to  
the option of calculating
\begin{equation}
\label{approx}
\br(\PH \to \PV\PV) \times \br(\PV \to \Pf\Pfb) 
                    \times \br(\PV \to \Pf\Pfb) 
                    \times \mbox{(statistical factor)}
\end{equation}
where $\PV = \PW, \PZ$.  The formula \refE{approx} is based on narrow
Higgs-width approximation and supposes the $\PW$ and $\PZ$ gauge
bosons to be on shell and thus neglects, in particular, all
interferences between different four-fermion final states. This
approximation is generally not accurate enough for Higgs masses below
and near the $\PW\PW/\PZ\PZ$ threshold.  A study of the ratio
of \refE{approx} over the \Prophecy\ prediction is reported and
discussed in \citere{BRTWiki}.

%%% Local Variables: 
%%% mode: latex
%%% TeX-master: ./LHCHiggsBRNote.tex
%%% End: 

%% file: comparison.tex
\section{Comparison with Previous Calculations}
\label{sec:Comparison}

%Comparison with  Djouadi-Baglio's results.
%Comparison with A. Djouadi, M. Spira, P.M. Zerwas hep-ph/9511344.

The results presented in the previous sections can be compared to
other estimates of the SM Higgs-boson BR uncertainties. An
early evaluation can be found in \citere{Djouadi:1995gt}, where
basically the same method has been applied to the parametric
uncertainties as the one used here, see \refS{sec:Procedure}.%
\footnote{The theoretical uncertainties have been omitted in
  \citere{Djouadi:1995gt}, while we did not include the
  parametric uncertainties due to the strange mass, since these only
  affect the $\Hss$ decays but have a negligible impact on all other
  branching ratios.} Differences in the size of the uncertainties
exist due to the improved experimental determinations in the quark
masses.
% and the extended existing higher-order calculations. 
At the
same time as \citere{Djouadi:1995gt} another work \cite{Gross:1994fv}
appeared which studied the parametric uncertainties of the SM Higgs
branching ratios, too. However, their treatment of the running
$\MSbar$ bottom and charm quark masses is not
consistent with the extractions of their values from QCD sum rules.
%$\MSbar$ quark masses for the charm and bottom quarks is not
%differs from the
%extractions of their values from QCD sum rules in an inconsistent way.

Very recently an analysis of the uncertainties in the SM Higgs-boson
BR calculations was published in \citere{Baglio:2010ae}.%
\footnote{The uncertainties on the decays $\Htt$ and $\Hgg$ have not
  been considered in detail in \citere{Baglio:2010ae} because they
  cannot be measured at the LHC.}  That analysis differs from ours in
various ways. Most importantly, \citere{Baglio:2010ae} uses the PDG
errors of $\Mb$ and $\Mc$ relative to their two-loop \MSbar\ value,
which exhibits a significant sensitivity to $\alphas$. In our
analysis, on the other hand we used the 1-loop pole masses of the
charm and bottom quarks which develop only a small dependence on the
strong coupling $\alphas$ \cite{narison}.  Moreover, we adopted
smaller uncertainties, as discussed in \refS{sec:Procedure}, which are
considered more realistic than the PDG values. We furthermore included
the parametric uncertainty on $\Mt$, which is relevant for large
values of $\MH$.  The values and uncertainties used in the two
analyses are summarized in Tab.~\ref{tab:inputcomp}.  Theory errors
due to missing higher-order corrections, see \refS{sec:Procedure}, are
included in our error analysis, but have been neglected in
\citere{Baglio:2010ae}.  Owing to the larger set of uncertainties
included, the analysis shown in \refS{sec:Results} should be
considered as more complete than the one presented in
\citere{Baglio:2010ae}.

The largest numerical differences between the uncertainties presented in
\refS{sec:Results} and in \citere{Baglio:2010ae} originate from the
different values of the quark-mass uncertainties, as shown in
Table~\ref{tab:inputcomp}. A direct comparison shows that the (more
appropriate) choice taken here, even taking into account the additional
sources of uncertainties, leads to a reduction of the total uncertainty
in $\br(\Hbb)$ by up to a factor of~3. For the channel
$\br(\HWW)$ a reduction of up to a factor of~4 can be observed. The
$\br(\Hcc)$ uncertainties are effectively lowered by a factor of~2, while
for $\br(\Hgg)$ the central values differ and slightly larger uncertainties
are observed.%
\footnote{The central values for $\br(\Hgg)$ in \citere{Baglio:2010ae}
  differ due to a bug in their implementation of HDECAY.} Especially
the substantially reduced uncertainty in $\br(\HWW)$ is crucial for
the correct interpretation of Tevatron and LHC Higgs search data
around $\MH = 2
\MW$~\cite{fermilab:2011,Aad:2011qi,Chatrchyan:2011tz,EPS11}. The
substantially smaller uncertainty in $\br(\Hbb)$ will be important for
the accurate interpretation of the LHC Higgs search data for $\MH \le
135 \UGeV$.

\begin{table}
\renewcommand{\arraystretch}{1.4}
\renewcommand{\tabcolsep}{3ex}
\begin{tabular}{lccc}
\hline 
{\bf Parameter} & {\bf this work} & {\bf \citere{Baglio:2010ae}} \\
\hline
$\alphas(\MZ)$ & 0.119  $\pm$ 0.002   & 0.1171  $\pm$ 0.0014 \\
$\MSbar$ mass $\Mb$ [GeV]& 4.16  $\pm$ 0.06  & 4.19$^{+0.18}_{-0.06}$ \\
$\MSbar$ mass $\Mc$ [GeV]& 1.28  $\pm$ 0.03  & 1.27$^{+0.07}_{-0.09}$ \\
pole mass $\Mt$ [GeV]& 172.5 $\pm$ 2.5 & 173.1 \\
\hline
\end{tabular}
\caption{Comparison of the uncertainties of the input parameter values
  used in this work and in \citere{Baglio:2010ae}. 
An uncertainty due to the top mass was not included in
\Bref{Baglio:2010ae}.} 
\label{tab:inputcomp}
\end{table}

%%% Local Variables: 
%%% mode: latex
%%% TeX-master: t
%%% End: 

%% file: conclusions.tex
\section{Summary and Conclusion}
\label{sec:Conclusion}

For the Higgs searches of the experimental collaborations at the LHC and
Tevatron precise predictions of the cross sections and decay branching
ratios are required. In this note we have presented updated numbers for
the total Higgs-boson decay width and all experimentally relevant
branching ratios. We have supplemented the predictions by uncertainties,
which have been estimated based on missing higher orders in the
theoretical calculations and from experimental uncertainties in the input
parameters. 
Specifically we took variations of the masses of the charm, bottom and
top quarks, as well as in the strong coupling constant into account.
Uncertainties of other parameters are irrelevant. As most of the error
sources are systematic or theoretical, the uncertainties should be
considered as conservative upper bounds rather than Gaussian errors.
Including an accurate estimate of these uncertainties is crucial for a correct
interpretation of LHC (and Tevatron) Higgs-boson searches.

%%% Local Variables: 
%%% mode: latex
%%% TeX-master: t
%%% End: 

%% file: appendix.tex
%\section{}
%\label{sec:AppendixA}

\begin{table}[htb]
\renewcommand{\arraystretch}{1.4}
\setlength{\tabcolsep}{3ex}
\begin{tabular}{lcccc}
\hline
$\MH$ [GeV] & $\Hbb$ & $\Htautau$ &$\Hmumu$ & $\Hcc$ \\
\hline
90.0   & 8.10E-01$^{+1.6\%}_{-1.6\%}$	   & 8.33E-02$^{+7.0\%}_{-6.8\%}$      & 2.89E-04$^{+7.3\%}_{-7.1\%}$	& 4.10E-02$^{+12.1\%}_{-12.1\%}$     \\
95.0   & 8.02E-01$^{+1.7\%}_{-1.7\%}$	   & 8.32E-02$^{+6.9\%}_{-6.8\%}$      & 2.89E-04$^{+7.3\%}_{-7.1\%}$	& 4.05E-02$^{+12.2\%}_{-12.1\%}$     \\
100.0  & 7.89E-01$^{+1.8\%}_{-1.8\%}$	   & 8.28E-02$^{+6.9\%}_{-6.7\%}$      & 2.88E-04$^{+7.2\%}_{-7.0\%}$	& 3.99E-02$^{+12.2\%}_{-12.2\%}$     \\
105.0  & 7.71E-01$^{+1.9\%}_{-2.0\%}$	   & 8.17E-02$^{+6.8\%}_{-6.7\%}$      & 2.84E-04$^{+7.1\%}_{-7.0\%}$	& 3.90E-02$^{+12.2\%}_{-12.2\%}$     \\
110.0  & 7.44E-01$^{+2.1\%}_{-2.2\%}$	   & 7.95E-02$^{+6.6\%}_{-6.5\%}$      & 2.76E-04$^{+7.0\%}_{-6.8\%}$	& 3.75E-02$^{+12.2\%}_{-12.2\%}$     \\
110.5  & 7.40E-01$^{+2.1\%}_{-2.2\%}$	   & 7.91E-02$^{+6.6\%}_{-6.5\%}$      & 2.75E-04$^{+6.9\%}_{-6.8\%}$	& 3.74E-02$^{+12.2\%}_{-12.2\%}$     \\
111.0  & 7.37E-01$^{+2.2\%}_{-2.2\%}$	   & 7.89E-02$^{+6.6\%}_{-6.5\%}$      & 2.74E-04$^{+6.9\%}_{-6.8\%}$	& 3.72E-02$^{+12.2\%}_{-12.2\%}$     \\
\hline
\end{tabular}
\caption{SM Higgs branching ratios to two fermions and their total uncertainties (expressed in percentage). Very low mass range.}
\label{BR-2fa}
\end{table}

\begin{table}[htb]
\renewcommand{\arraystretch}{1.4}
\setlength{\tabcolsep}{3ex}
\begin{tabular}{lcccc}
\hline
$\MH$ [GeV] & $\Hbb$ & $\Htautau$ &$\Hmumu$ & $\Hcc$ \\
\hline
130.5  & 4.85E-01$^{+3.8\%}_{-3.8\%}$	 & 5.36E-02$^{+5.2\%}_{-5.2\%}$    & 1.86E-04$^{+5.5\%}_{-5.4\%}$    & 2.45E-02$^{+12.2\%}_{-12.1\%}$	  \\
131.0  & 4.76E-01$^{+3.8\%}_{-3.9\%}$	 & 5.26E-02$^{+5.2\%}_{-5.1\%}$    & 1.82E-04$^{+5.4\%}_{-5.3\%}$    & 2.41E-02$^{+12.2\%}_{-12.1\%}$	  \\
131.5  & 4.67E-01$^{+3.9\%}_{-3.9\%}$	 & 5.17E-02$^{+5.1\%}_{-5.1\%}$    & 1.80E-04$^{+5.4\%}_{-5.3\%}$    & 2.36E-02$^{+12.2\%}_{-12.1\%}$	  \\
132.0  & 4.58E-01$^{+3.9\%}_{-4.0\%}$	 & 5.07E-02$^{+5.1\%}_{-5.0\%}$    & 1.77E-04$^{+5.3\%}_{-5.2\%}$    & 2.32E-02$^{+12.2\%}_{-12.1\%}$	  \\
132.5  & 4.49E-01$^{+4.0\%}_{-4.0\%}$	 & 4.97E-02$^{+5.0\%}_{-5.0\%}$    & 1.72E-04$^{+5.2\%}_{-5.2\%}$    & 2.27E-02$^{+12.2\%}_{-12.1\%}$	  \\
133.0  & 4.40E-01$^{+4.0\%}_{-4.1\%}$	 & 4.88E-02$^{+5.0\%}_{-4.9\%}$    & 1.69E-04$^{+5.2\%}_{-5.1\%}$    & 2.23E-02$^{+12.2\%}_{-12.1\%}$	  \\
133.5  & 4.31E-01$^{+4.1\%}_{-4.2\%}$	 & 4.79E-02$^{+4.9\%}_{-4.9\%}$    & 1.67E-04$^{+5.1\%}_{-5.1\%}$    & 2.18E-02$^{+12.2\%}_{-12.1\%}$	  \\
\hline
\end{tabular}
\caption{SM Higgs branching ratios to two fermions and their total uncertainties (expressed in percentage). Low and
intermediate mass range.}
\label{BR-2fb}
\end{table}

\begin{table}[htb]
\renewcommand{\arraystretch}{1.4}
\setlength{\tabcolsep}{3ex}
\begin{tabular}{lcccc}
\hline
$\MH$ [GeV] & $\Hbb$ & $\Htautau$ &$\Hmumu$ & $\Hcc$ \\
\hline
172.0  & 7.10E-03$^{+4.5\%}_{-4.5\%}$	  & 8.25E-04$^{+2.6\%}_{-2.6\%}$      & 2.86E-06$^{+2.6\%}_{-2.6\%}$	 & 3.57E-04$^{+9.9\%}_{-9.9\%}$       \\
174.0  & 6.45E-03$^{+4.5\%}_{-4.5\%}$	  & 7.51E-04$^{+2.6\%}_{-2.6\%}$      & 2.60E-06$^{+2.6\%}_{-2.6\%}$	 & 3.25E-04$^{+10.0\%}_{-9.9\%}$       \\
176.0  & 5.91E-03$^{+4.6\%}_{-4.5\%}$	  & 6.89E-04$^{+2.6\%}_{-2.6\%}$      & 2.39E-06$^{+2.7\%}_{-2.6\%}$	 & 2.97E-04$^{+10.0\%}_{-9.9\%}$       \\
178.0  & 5.44E-03$^{+4.6\%}_{-4.5\%}$	  & 6.36E-04$^{+2.6\%}_{-2.6\%}$      & 2.20E-06$^{+2.7\%}_{-2.6\%}$	 & 2.74E-04$^{+10.0\%}_{-9.9\%}$       \\
180.0  & 5.01E-03$^{+4.6\%}_{-4.5\%}$	  & 5.87E-04$^{+2.6\%}_{-2.6\%}$      & 2.04E-06$^{+2.7\%}_{-2.6\%}$	 & 2.52E-04$^{+10.0\%}_{-9.9\%}$       \\
182.0  & 4.55E-03$^{+4.6\%}_{-4.5\%}$	  & 5.34E-04$^{+2.6\%}_{-2.6\%}$      & 1.85E-06$^{+2.7\%}_{-2.6\%}$	 & 2.29E-04$^{+10.0\%}_{-9.9\%}$       \\
184.0  & 4.09E-03$^{+4.6\%}_{-4.5\%}$	  & 4.81E-04$^{+2.6\%}_{-2.6\%}$      & 1.67E-06$^{+2.7\%}_{-2.6\%}$	 & 2.06E-04$^{+10.0\%}_{-9.9\%}$       \\
\hline
\end{tabular}
\caption{SM Higgs branching ratios to two fermions and their total uncertainties (expressed in percentage). Intermediate mass range.}
\label{BR-2fc}
\end{table}

\begin{table}[htb]
\renewcommand{\arraystretch}{1.4}
\setlength{\tabcolsep}{2ex}
\begin{tabular}{lccccc}
\hline
$\MH$ [GeV] & $\Hbb$ & $\Htautau$ &$\Hmumu$ & $\Hcc$ & $\Htt$\\
\hline
262.0  & 8.74E-04$^{+4.7\%}_{-4.6\%}$	& 1.10E-04$^{+2.6\%}_{-2.6\%}$      & 3.81E-07$^{+2.6\%}_{-2.6\%}$    & 4.39E-05$^{+10.1\%}_{-10.0\%}$    & 1.89E-07$^{+342.0\%}_{-102.1\%}$	 \\
264.0  & 8.52E-04$^{+4.7\%}_{-4.6\%}$	& 1.07E-04$^{+2.6\%}_{-2.6\%}$      & 3.72E-07$^{+2.6\%}_{-2.6\%}$    & 4.28E-05$^{+10.1\%}_{-10.0\%}$    & 4.49E-07$^{+276.1\%}_{-91.5\%}$	 \\
266.0  & 8.31E-04$^{+4.7\%}_{-4.6\%}$	& 1.05E-04$^{+2.6\%}_{-2.6\%}$      & 3.63E-07$^{+2.6\%}_{-2.6\%}$    & 4.18E-05$^{+10.1\%}_{-10.0\%}$    & 8.63E-07$^{+210.1\%}_{-80.9\%}$	 \\
268.0  & 8.11E-04$^{+4.7\%}_{-4.6\%}$	& 1.02E-04$^{+2.6\%}_{-2.6\%}$      & 3.55E-07$^{+2.6\%}_{-2.6\%}$    & 4.08E-05$^{+10.1\%}_{-10.0\%}$    & 1.46E-06$^{+144.1\%}_{-70.3\%}$	 \\
270.0  & 7.92E-04$^{+4.7\%}_{-4.6\%}$	& 1.00E-04$^{+2.6\%}_{-2.6\%}$      & 3.47E-07$^{+2.6\%}_{-2.6\%}$    & 3.98E-05$^{+10.1\%}_{-10.0\%}$    & 2.29E-06$^{+78.1\%}_{-59.7\%}$	\\
272.0  & 7.73E-04$^{+4.7\%}_{-4.6\%}$	& 9.80E-05$^{+2.6\%}_{-2.6\%}$      & 3.40E-07$^{+2.6\%}_{-2.6\%}$    & 3.89E-05$^{+10.1\%}_{-10.0\%}$    & 3.37E-06$^{+72.3\%}_{-56.6\%}$	\\
274.0  & 7.55E-04$^{+4.7\%}_{-4.6\%}$	& 9.58E-05$^{+2.6\%}_{-2.6\%}$      & 3.32E-07$^{+2.6\%}_{-2.6\%}$    & 3.80E-05$^{+10.1\%}_{-10.0\%}$    & 4.72E-06$^{+66.5\%}_{-53.6\%}$	\\
\hline
\end{tabular}
\caption{SM Higgs branching ratios to two fermions and their total uncertainties (expressed in percentage). High mass range.}
\label{BR-2fd}
\end{table}

\begin{table}[htb]
\renewcommand{\arraystretch}{1.4}
\setlength{\tabcolsep}{2ex}
\begin{tabular}{lccccc}
\hline
$\MH$ [GeV] & $\Hbb$ & $\Htautau$ &$\Hmumu$ & $\Hcc$ & $\Htt$\\
\hline
550.0  & 8.64E-05$^{+7.7\%}_{-5.7\%}$	    & 1.23E-05$^{+5.5\%}_{-3.5\%}$	 & 4.27E-08$^{+5.5\%}_{-3.5\%}$       & 4.33E-06$^{+13.1\%}_{-11.1\%}$  	   & 1.83E-01$^{+5.4\%}_{-10.9\%}$	\\
560.0  & 8.28E-05$^{+7.8\%}_{-5.8\%}$	    & 1.19E-05$^{+5.6\%}_{-3.7\%}$	 & 4.10E-08$^{+5.6\%}_{-3.6\%}$       & 4.15E-06$^{+13.2\%}_{-11.2\%}$  	   & 1.80E-01$^{+5.6\%}_{-11.3\%}$	\\
570.0  & 7.95E-05$^{+7.8\%}_{-5.9\%}$	    & 1.14E-05$^{+5.8\%}_{-3.7\%}$	 & 3.95E-08$^{+5.7\%}_{-3.8\%}$       & 3.98E-06$^{+13.3\%}_{-11.3\%}$  	   & 1.78E-01$^{+5.7\%}_{-11.6\%}$	\\
580.0  & 7.64E-05$^{+7.9\%}_{-6.0\%}$	    & 1.10E-05$^{+5.8\%}_{-4.0\%}$	 & 3.80E-08$^{+5.8\%}_{-4.0\%}$       & 3.83E-06$^{+13.4\%}_{-11.4\%}$  	   & 1.75E-01$^{+5.9\%}_{-11.9\%}$	\\
590.0  & 7.34E-05$^{+8.1\%}_{-6.1\%}$	    & 1.06E-05$^{+6.0\%}_{-4.0\%}$	 & 3.67E-08$^{+6.0\%}_{-4.0\%}$       & 3.68E-06$^{+13.5\%}_{-11.5\%}$  	   & 1.72E-01$^{+6.2\%}_{-12.2\%}$	\\
600.0  & 7.06E-05$^{+8.2\%}_{-6.3\%}$	    & 1.02E-05$^{+6.2\%}_{-4.2\%}$	 & 3.54E-08$^{+6.1\%}_{-4.2\%}$       & 3.54E-06$^{+13.7\%}_{-11.6\%}$  	   & 1.69E-01$^{+6.3\%}_{-12.4\%}$	\\
610.0  & 6.81E-05$^{+8.3\%}_{-6.5\%}$	    & 9.87E-06$^{+6.2\%}_{-4.4\%}$	 & 3.42E-08$^{+6.2\%}_{-4.4\%}$       & 3.41E-06$^{+13.8\%}_{-11.8\%}$  	   & 1.66E-01$^{+6.4\%}_{-12.8\%}$	\\
\hline
\end{tabular}
\caption{SM Higgs branching ratios to two fermions and their total uncertainties (expressed in percentage). Very high mass range.}
\label{BR-2fe}
\end{table}

\begin{table}[htb]
\renewcommand{\arraystretch}{1.4}
\setlength{\tabcolsep}{1ex}
\begin{tabular}{lccccc|c}
\hline
$\MH$ [GeV] & $\Hgg$ & $\Hgaga$ &$\HZga$ & $\HWW$ & $\HZZ$ & $\Gamma_{\PH}$ [GeV]\\
\hline
90.0   & 6.12E-02$^{+11.8\%}_{-11.4\%}$      & 1.22E-03$^{+6.2\%}_{-6.2\%}$    & 0.00E+00$^{+0.0\%}_{-0.0\%}$	   & 2.07E-03$^{+5.8\%}_{-5.7\%}$      & 4.17E-04$^{+5.8\%}_{-5.6\%}$		   & 2.22E-03$^{+5.2\%}_{-5.2\%}$	\\
95.0   & 6.74E-02$^{+11.7\%}_{-11.3\%}$      & 1.39E-03$^{+6.2\%}_{-6.1\%}$    & 4.48E-06$^{+10.2\%}_{-10.1\%}$    & 4.67E-03$^{+5.8\%}_{-5.6\%}$      & 6.65E-04$^{+5.7\%}_{-5.6\%}$		   & 2.35E-03$^{+5.2\%}_{-5.1\%}$	\\
100.0  & 7.36E-02$^{+11.6\%}_{-11.2\%}$      & 1.58E-03$^{+6.2\%}_{-6.1\%}$    & 4.93E-05$^{+10.2\%}_{-10.1\%}$    & 1.09E-02$^{+5.7\%}_{-5.6\%}$      & 1.12E-03$^{+5.7\%}_{-5.5\%}$		   & 2.49E-03$^{+5.1\%}_{-5.1\%}$	\\
105.0  & 7.95E-02$^{+11.4\%}_{-11.1\%}$      & 1.77E-03$^{+6.1\%}_{-5.9\%}$    & 1.71E-04$^{+10.2\%}_{-9.9\%}$     & 2.40E-02$^{+5.6\%}_{-5.5\%}$      & 2.13E-03$^{+5.6\%}_{-5.5\%}$		   & 2.65E-03$^{+5.1\%}_{-5.0\%}$	\\
110.0  & 8.45E-02$^{+11.2\%}_{-10.9\%}$      & 1.95E-03$^{+5.9\%}_{-5.8\%}$    & 3.91E-04$^{+9.9\%}_{-9.8\%}$	   & 4.77E-02$^{+5.4\%}_{-5.3\%}$      & 4.34E-03$^{+5.4\%}_{-5.3\%}$		   & 2.85E-03$^{+4.9\%}_{-4.9\%}$	\\
110.5  & 8.49E-02$^{+11.2\%}_{-10.8\%}$      & 1.97E-03$^{+5.9\%}_{-5.8\%}$    & 4.18E-04$^{+9.9\%}_{-9.8\%}$	   & 5.09E-02$^{+5.4\%}_{-5.3\%}$      & 4.67E-03$^{+5.4\%}_{-5.3\%}$		   & 2.87E-03$^{+4.9\%}_{-4.9\%}$	\\
111.0  & 8.53E-02$^{+11.2\%}_{-10.8\%}$      & 1.98E-03$^{+5.9\%}_{-5.8\%}$    & 4.47E-04$^{+9.9\%}_{-9.8\%}$	   & 5.40E-02$^{+5.4\%}_{-5.2\%}$      & 5.00E-03$^{+5.4\%}_{-5.2\%}$		   & 2.90E-03$^{+4.9\%}_{-4.8\%}$	\\
\hline
\end{tabular}
\caption{SM Higgs branching ratios to two gauge bosons and Higgs total width together with their total uncertainties (expressed in percentage). Very low mass range.}
\label{BR-2gba}
\end{table}

\begin{table}[htb]
\renewcommand{\arraystretch}{1.4}
\setlength{\tabcolsep}{1ex}
\begin{tabular}{lccccc|c}
\hline
$\MH$ [GeV] & $\Hgg$ & $\Hgaga$ &$\HZga$ & $\HWW$ & $\HZZ$ & $\Gamma_{\PH}$ [GeV]\\
\hline
130.5  & 7.90E-02$^{+9.8\%}_{-9.5\%}$	    & 2.24E-03$^{+4.5\%}_{-4.3\%}$	 & 1.99E-03$^{+8.4\%}_{-8.4\%}$      & 3.12E-01$^{+3.6\%}_{-3.5\%}$     & 4.13E-02$^{+3.6\%}_{-3.5\%}$     & 5.01E-03$^{+3.4\%}_{-3.4\%}$	\\
131.0  & 7.82E-02$^{+9.7\%}_{-9.5\%}$	    & 2.23E-03$^{+4.4\%}_{-4.3\%}$	 & 2.01E-03$^{+8.3\%}_{-8.3\%}$      & 3.22E-01$^{+3.5\%}_{-3.5\%}$     & 4.28E-02$^{+3.5\%}_{-3.5\%}$     & 5.12E-03$^{+3.4\%}_{-3.3\%}$	\\
131.5  & 7.73E-02$^{+9.7\%}_{-9.4\%}$	    & 2.22E-03$^{+4.4\%}_{-4.2\%}$	 & 2.06E-03$^{+8.3\%}_{-8.3\%}$      & 3.31E-01$^{+3.5\%}_{-3.4\%}$     & 4.42E-02$^{+3.5\%}_{-3.4\%}$     & 5.23E-03$^{+3.3\%}_{-3.3\%}$	\\
132.0  & 7.65E-02$^{+9.6\%}_{-9.4\%}$	    & 2.21E-03$^{+4.3\%}_{-4.2\%}$	 & 2.09E-03$^{+8.2\%}_{-8.2\%}$      & 3.41E-01$^{+3.4\%}_{-3.4\%}$     & 4.57E-02$^{+3.4\%}_{-3.4\%}$     & 5.35E-03$^{+3.2\%}_{-3.2\%}$	\\
132.5  & 7.56E-02$^{+9.6\%}_{-9.4\%}$	    & 2.20E-03$^{+4.2\%}_{-4.1\%}$	 & 2.13E-03$^{+8.2\%}_{-8.2\%}$      & 3.51E-01$^{+3.3\%}_{-3.3\%}$     & 4.72E-02$^{+3.3\%}_{-3.3\%}$     & 5.47E-03$^{+3.2\%}_{-3.2\%}$	\\
133.0  & 7.47E-02$^{+9.5\%}_{-9.3\%}$	    & 2.19E-03$^{+4.2\%}_{-4.1\%}$	 & 2.16E-03$^{+8.1\%}_{-8.1\%}$      & 3.61E-01$^{+3.3\%}_{-3.2\%}$     & 4.87E-02$^{+3.3\%}_{-3.2\%}$     & 5.60E-03$^{+3.1\%}_{-3.1\%}$	\\
133.5  & 7.37E-02$^{+9.5\%}_{-9.3\%}$	    & 2.17E-03$^{+4.1\%}_{-4.0\%}$	 & 2.19E-03$^{+8.1\%}_{-8.1\%}$      & 3.70E-01$^{+3.2\%}_{-3.2\%}$     & 5.02E-02$^{+3.2\%}_{-3.2\%}$     & 5.74E-03$^{+3.1\%}_{-3.1\%}$	\\
\hline
\end{tabular}
\caption{SM Higgs branching ratios to two gauge bosons and Higgs total width together with their total uncertainties (expressed in percentage). Low and intermediate mass range.}
\label{BR-2gbb}
\end{table}

\begin{table}[htb]
\renewcommand{\arraystretch}{1.4}
\setlength{\tabcolsep}{1ex}
\begin{tabular}{lccccc|c}
\hline
$\MH$ [GeV] & $\Hgg$ & $\Hgaga$ &$\HZga$ & $\HWW$ & $\HZZ$ & $\Gamma_{\PH}$ [GeV]\\
\hline
172.0  & 2.01E-03$^{+7.7\%}_{-7.6\%}$	 & 1.43E-04$^{+1.6\%}_{-1.6\%}$    & 3.70E-04$^{+5.6\%}_{-5.6\%}$    & 9.63E-01$^{+0.0\%}_{-0.0\%}$	  & 2.61E-02$^{+0.1\%}_{-0.1\%}$      & 4.29E-01$^{+0.6\%}_{-0.6\%}$	   \\
174.0  & 1.88E-03$^{+7.7\%}_{-7.6\%}$	 & 1.32E-04$^{+1.6\%}_{-1.6\%}$    & 3.48E-04$^{+5.6\%}_{-5.5\%}$    & 9.60E-01$^{+0.0\%}_{-0.0\%}$	  & 2.98E-02$^{+0.1\%}_{-0.1\%}$      & 4.77E-01$^{+0.6\%}_{-0.6\%}$	   \\
176.0  & 1.76E-03$^{+7.7\%}_{-7.6\%}$	 & 1.22E-04$^{+1.6\%}_{-1.6\%}$    & 3.29E-04$^{+5.6\%}_{-5.5\%}$    & 9.55E-01$^{+0.0\%}_{-0.0\%}$	  & 3.54E-02$^{+0.1\%}_{-0.1\%}$      & 5.25E-01$^{+0.6\%}_{-0.6\%}$	   \\
178.0  & 1.65E-03$^{+7.7\%}_{-7.6\%}$	 & 1.14E-04$^{+1.6\%}_{-1.6\%}$    & 3.12E-04$^{+5.6\%}_{-5.5\%}$    & 9.47E-01$^{+0.0\%}_{-0.0\%}$	  & 4.44E-02$^{+0.1\%}_{-0.1\%}$      & 5.75E-01$^{+0.6\%}_{-0.6\%}$	   \\
180.0  & 1.56E-03$^{+7.7\%}_{-7.6\%}$	 & 1.05E-04$^{+1.6\%}_{-1.6\%}$    & 2.96E-04$^{+5.6\%}_{-5.5\%}$    & 9.32E-01$^{+0.0\%}_{-0.0\%}$	  & 6.02E-02$^{+0.1\%}_{-0.1\%}$      & 6.31E-01$^{+0.6\%}_{-0.6\%}$	   \\
182.0  & 1.44E-03$^{+7.7\%}_{-7.6\%}$	 & 9.69E-05$^{+1.6\%}_{-1.6\%}$    & 2.76E-04$^{+5.6\%}_{-5.5\%}$    & 9.03E-01$^{+0.0\%}_{-0.0\%}$	  & 9.00E-02$^{+0.1\%}_{-0.1\%}$      & 7.00E-01$^{+0.6\%}_{-0.6\%}$	   \\
184.0  & 1.32E-03$^{+7.7\%}_{-7.6\%}$	 & 8.81E-05$^{+1.6\%}_{-1.6\%}$    & 2.54E-04$^{+5.6\%}_{-5.5\%}$    & 8.62E-01$^{+0.0\%}_{-0.0\%}$	  & 1.31E-01$^{+0.0\%}_{-0.0\%}$      & 7.88E-01$^{+0.6\%}_{-0.6\%}$	   \\
\hline
\end{tabular}
\caption{SM Higgs branching ratios to two gauge bosons and Higgs total width together with their their total uncertainties (expressed in percentage). Intermediate mass range.}
\label{BR-2gbc}
\end{table}

\begin{table}[htb]
\renewcommand{\arraystretch}{1.4}
\setlength{\tabcolsep}{1ex}
\begin{tabular}{lccccc|c}
\hline
$\MH$ [GeV] & $\Hgg$ & $\Hgaga$ &$\HZga$ & $\HWW$ & $\HZZ$ & $\Gamma_{\PH}$ [GeV]\\
\hline
262.0  & 6.64E-04$^{+7.7\%}_{-7.5\%}$	     & 2.11E-05$^{+1.7\%}_{-1.7\%}$	& 8.39E-05$^{+5.6\%}_{-5.6\%}$  	    & 6.98E-01$^{+0.0\%}_{-0.0\%}$		& 3.00E-01$^{+0.0\%}_{-0.0\%}$  	    & 4.91E+00$^{+0.5\%}_{-0.5\%}$	 \\
264.0  & 6.61E-04$^{+7.7\%}_{-7.5\%}$	     & 2.06E-05$^{+1.7\%}_{-1.7\%}$	& 8.21E-05$^{+5.6\%}_{-5.6\%}$  	    & 6.98E-01$^{+0.0\%}_{-0.0\%}$		& 3.00E-01$^{+0.0\%}_{-0.0\%}$  	    & 5.07E+00$^{+0.5\%}_{-0.5\%}$	 \\
266.0  & 6.59E-04$^{+7.7\%}_{-7.6\%}$	     & 2.01E-05$^{+1.7\%}_{-1.7\%}$	& 8.04E-05$^{+5.6\%}_{-5.6\%}$  	    & 6.98E-01$^{+0.0\%}_{-0.0\%}$		& 3.01E-01$^{+0.0\%}_{-0.0\%}$  	    & 5.23E+00$^{+0.5\%}_{-0.5\%}$	 \\
268.0  & 6.57E-04$^{+7.7\%}_{-7.6\%}$	     & 1.96E-05$^{+1.7\%}_{-1.7\%}$	& 7.88E-05$^{+5.6\%}_{-5.6\%}$  	    & 6.97E-01$^{+0.0\%}_{-0.0\%}$		& 3.01E-01$^{+0.0\%}_{-0.0\%}$  	    & 5.39E+00$^{+0.5\%}_{-0.5\%}$	 \\
270.0  & 6.55E-04$^{+7.7\%}_{-7.6\%}$	     & 1.91E-05$^{+1.7\%}_{-1.7\%}$	& 7.72E-05$^{+5.6\%}_{-5.6\%}$  	    & 6.97E-01$^{+0.0\%}_{-0.0\%}$		& 3.02E-01$^{+0.0\%}_{-0.0\%}$  	    & 5.55E+00$^{+0.5\%}_{-0.5\%}$	 \\
272.0  & 6.54E-04$^{+7.7\%}_{-7.6\%}$	     & 1.87E-05$^{+1.7\%}_{-1.7\%}$	& 7.56E-05$^{+5.6\%}_{-5.6\%}$  	    & 6.96E-01$^{+0.0\%}_{-0.0\%}$		& 3.02E-01$^{+0.0\%}_{-0.0\%}$  	    & 5.72E+00$^{+0.5\%}_{-0.5\%}$	 \\
274.0  & 6.53E-04$^{+7.7\%}_{-7.6\%}$	     & 1.82E-05$^{+1.7\%}_{-1.7\%}$	& 7.41E-05$^{+5.6\%}_{-5.6\%}$  	    & 6.96E-01$^{+0.0\%}_{-0.0\%}$		& 3.02E-01$^{+0.0\%}_{-0.0\%}$  	    & 5.89E+00$^{+0.5\%}_{-0.5\%}$	 \\
\hline
\end{tabular}
\caption{SM Higgs branching ratios to two gauge bosons and Higgs total width together with their their total uncertainties (expressed in percentage). High mass range.}
\label{BR-2gbd}
\end{table}

\begin{table}[htb]
\renewcommand{\arraystretch}{1.4}
\setlength{\tabcolsep}{1ex}
\begin{tabular}{lccccc|c}
\hline
$\MH$ [GeV] & $\Hgg$ & $\Hgaga$ &$\HZga$ & $\HWW$ & $\HZZ$ & $\Gamma_{\PH}$ [GeV]\\
\hline
550.0  & 5.05E-04$^{+10.4\%}_{-8.4\%}$       & 9.90E-08$^{+6.1\%} _{-2.7\%}$	  & 5.30E-06$^{+8.6\%}_{-6.6\%}$      & 5.50E-01$^{+2.4\%}_{-1.2\%}$    & 2.66E-01$^{+2.5\%}_{-1.2\%}$      & 9.30E+01$^{+1.6\%}_{-3.5\%}$	 \\
560.0  & 4.88E-04$^{+10.5\%}_{-8.6\%}$       & 8.10E-08$^{+8.7\%} _{-3.2\%}$	  & 4.95E-06$^{+8.6\%}_{-6.6\%}$      & 5.52E-01$^{+2.5\%}_{-1.2\%}$    & 2.67E-01$^{+2.5\%}_{-1.2\%}$      & 9.86E+01$^{+1.7\%}_{-3.6\%}$	 \\
570.0  & 4.71E-04$^{+10.6\%}_{-8.7\%}$       & 6.83E-08$^{+11.5\%}_{-7.2\%}$	  & 4.64E-06$^{+8.6\%}_{-6.7\%}$      & 5.53E-01$^{+2.5\%}_{-1.2\%}$    & 2.68E-01$^{+2.5\%}_{-1.3\%}$      & 1.04E+02$^{+1.9\%}_{-3.7\%}$	 \\
580.0  & 4.55E-04$^{+10.7\%}_{-8.9\%}$       & 5.91E-08$^{+15.2\%}_{-10.3\%}$	  & 4.35E-06$^{+8.7\%}_{-6.9\%}$      & 5.55E-01$^{+2.5\%}_{-1.2\%}$    & 2.70E-01$^{+2.5\%}_{-1.3\%}$      & 1.10E+02$^{+2.0\%}_{-3.8\%}$	 \\
590.0  & 4.39E-04$^{+10.9\%}_{-8.9\%}$       & 5.33E-08$^{+17.8\%}_{-13.5\%}$	  & 4.08E-06$^{+8.9\%}_{-6.9\%}$      & 5.57E-01$^{+2.5\%}_{-1.3\%}$    & 2.71E-01$^{+2.5\%}_{-1.3\%}$      & 1.16E+02$^{+2.2\%}_{-3.9\%}$	 \\
600.0  & 4.24E-04$^{+11.1\%}_{-9.1\%}$       & 5.03E-08$^{+20.7\%}_{-16.0\%}$	  & 3.84E-06$^{+9.2\%}_{-7.1\%}$      & 5.58E-01$^{+2.5\%}_{-1.3\%}$    & 2.72E-01$^{+2.5\%}_{-1.3\%}$      & 1.23E+02$^{+2.3\%}_{-4.0\%}$	 \\
610.0  & 4.10E-04$^{+11.0\%}_{-9.3\%}$       & 4.86E-08$^{+22.0\%}_{-17.2\%}$	  & 3.61E-06$^{+9.3\%}_{-7.5\%}$      & 5.60E-01$^{+2.5\%}_{-1.3\%}$    & 2.74E-01$^{+2.6\%}_{-1.3\%}$      & 1.29E+02$^{+2.5\%}_{-4.2\%}$	 \\
\hline
\end{tabular}
\caption{SM Higgs branching ratios to two gauge bosons and Higgs total width together with their their total uncertainties (expressed in percentage). Very high mass range.}
\label{BR-2gbe}
\end{table}

%\newpage
%\section{Higgs Decays into four fermions}
%\label{sec:AppendixB}

\begin{table}[htb]
\renewcommand{\arraystretch}{1.3}
\setlength{\tabcolsep}{1.5ex}
\begin{tabular}{lccccccccc}
\hline
$\MH$  & $\PH \to 4\Pl$ & $\PH \to 4\Pl$ & $\PH \to \Pe\Pe\Pe\Pe$ & $\PH \to \Pe\Pe\mu\mu$ &
$\PH \to 2\Pl2\nu$ & $\PH \to 2\Pl2\nu$ & $\PH \to \Pe\nu \Pe \nu$ & $\PH \to \Pe \nu \mu \nu$ &$\Delta$BR  \\
%$[\UGeV]$ & $\Pl=\Pe,\mu,\tau$ & $\Pl=\Pe,\mu$ & & & $\Pl=\Pe,\mu,\tau$ - $\nu={}$any & $\Pl=\Pe,\mu$ - $\nu={}$any& & & [\%]\\ 
$[\UGeV]$ & $\Pl=\Pe,\mu,\tau$ & $\Pl=\Pe,\mu$ & & & $\Pl=\Pe,\mu,\tau$ & $\Pl=\Pe,\mu$ & & & [\%]\\ 
 & & & & & $\nu={}$any & $\nu={}$any& & & \\ 
\hline
90.0 & 4.89E-06 & 2.33E-06 & 7.01E-07 & 9.30E-07 & 2.10E-04 & 9.12E-05 & 1.75E-05 & 2.43E-05  & $\pm$5.8\\
95.0 & 7.73E-06 & 3.67E-06 & 1.10E-06 & 1.48E-06 & 4.80E-04 & 2.10E-04 & 4.42E-05 & 5.49E-05  & $\pm$5.8\\
100.0 & 1.28E-05 & 6.06E-06 & 1.78E-06 & 2.49E-06 & 1.14E-03 & 5.03E-04 & 1.13E-04 & 1.29E-04 & $\pm$5.7\\
105.0 & 2.37E-05 & 1.11E-05 & 3.18E-06 & 4.74E-06 & 2.53E-03 & 1.12E-03 & 2.59E-04 & 2.83E-04 & $\pm$5.6\\
110.0 & 4.72E-05 & 2.18E-05 & 6.04E-06 & 9.69E-06 & 5.08E-03 & 2.26E-03 & 5.32E-04 & 5.61E-04 & $\pm$5.4\\
\hline
\end{tabular}
\caption{SM Higgs branching ratios for the different $\PH \to 4\Pl$ and
$\PH \to 2\Pl2\nu$ final states and their total uncertainties (expressed in percentage). Very low mass range.}
\label{BR-4f1a}
\end{table}

\begin{table}[htb]
\renewcommand{\arraystretch}{1.3}
\setlength{\tabcolsep}{1.5ex}
\begin{tabular}{lccccccccc}
\hline
$\MH$  & $\PH \to 4\Pl$ & $\PH \to 4\Pl$ & $\PH \to \Pe\Pe\Pe\Pe$ & $\PH \to \Pe\Pe\mu\mu$ &
$\PH \to 2\Pl2\nu$ & $\PH \to 2\Pl2\nu$ & $\PH \to\Pe\nu\Pe  \nu$ & $\PH \to\Pe\nu \mu \nu$ &$\Delta$BR  \\
%$[\UGeV]$ & $\Pl=\Pe,\mu,\tau$ & $\Pl=\Pe,\mu$ & & & $\Pl=\Pe,\mu,\tau$ - $\nu={}$any & $\Pl=\Pe,\mu$ - $\nu={}$any& & & [\%]\\ 
$[\UGeV]$ & $\Pl=\Pe,\mu,\tau$ & $\Pl=\Pe,\mu$ & & & $\Pl=\Pe,\mu,\tau$ & $\Pl=\Pe,\mu$ & & & [\%]\\ 
 & & & & & $\nu={}$any & $\nu={}$any& & & \\ 
\hline
130.5 & 4.29E-04 & 1.93E-04 & 5.02E-05 & 9.29E-05 & 3.43E-02 & 1.55E-02 & 3.71E-03 & 3.67E-03 & $\pm$3.6\\
131.0 & 4.44E-04 & 2.00E-04 & 5.19E-05 & 9.62E-05 & 3.54E-02 & 1.60E-02 & 3.83E-03 & 3.79E-03 & $\pm$3.5\\
131.5 & 4.59E-04 & 2.07E-04 & 5.37E-05 & 9.95E-05 & 3.64E-02 & 1.65E-02 & 3.94E-03 & 3.90E-03 & $\pm$3.5\\
132.0 & 4.75E-04 & 2.14E-04 & 5.54E-05 & 1.03E-04 & 3.75E-02 & 1.70E-02 & 4.06E-03 & 4.01E-03 & $\pm$3.4\\
132.5 & 4.90E-04 & 2.21E-04 & 5.71E-05 & 1.06E-04 & 3.86E-02 & 1.75E-02 & 4.18E-03 & 4.13E-03 & $\pm$3.3\\
\hline
\end{tabular}
\caption{SM Higgs branching ratios for the different $\PH \to 4\Pl$ and
$\PH \to 2\Pl2\nu$ final states and their total uncertainties (expressed in percentage). Low and intermediate mass range.}
\label{BR-4f1b}
\end{table}

\begin{table}[htb]
\setlength{\tabcolsep}{1.5ex}
\renewcommand{\arraystretch}{1.3}
\begin{tabular}{lccccccccc}
\hline
$\MH$  & $\PH \to 4\Pl$ & $\PH \to 4\Pl$ & $\PH \to \Pe\Pe\Pe\Pe$ & $\PH \to \Pe\Pe\mu\mu$ &
$\PH \to 2\Pl2\nu$ & $\PH \to 2\Pl2\nu$ & $\PH \to\Pe\nu\Pe \nu$ & $\PH \to\Pe\nu \mu \nu$ &$\Delta$BR  \\
%$[\UGeV]$ & $\Pl=\Pe,\mu,\tau$ & $\Pl=\Pe,\mu$ & & & $\Pl=\Pe,\mu,\tau$ - $\nu={}$any & $\Pl=\Pe,\mu$ - $\nu={}$any& & & [\%]\\ 
$[\UGeV]$ & $\Pl=\Pe,\mu,\tau$ & $\Pl=\Pe,\mu$ & & & $\Pl=\Pe,\mu,\tau$ & $\Pl=\Pe,\mu$ & & & [\%]\\ 
 & & & & & $\nu={}$any & $\nu={}$any& & & \\ 
\hline
172.0 & 2.66E-04 & 1.19E-04 & 3.00E-05 & 5.87E-05 & 1.03E-01 & 4.61E-02 & 1.15E-02 & 1.14E-02 & $\pm$0.0\\
174.0 & 3.04E-04 & 1.36E-04 & 3.42E-05 & 6.71E-05 & 1.03E-01 & 4.61E-02 & 1.14E-02 & 1.13E-02 & $\pm$0.0\\
176.0 & 3.61E-04 & 1.61E-04 & 4.06E-05 & 7.98E-05 & 1.03E-01 & 4.60E-02 & 1.14E-02 & 1.13E-02 & $\pm$0.0\\
178.0 & 4.52E-04 & 2.01E-04 & 5.07E-05 & 9.99E-05 & 1.02E-01 & 4.58E-02 & 1.13E-02 & 1.12E-02 & $\pm$0.0\\
180.0 & 6.12E-04 & 2.73E-04 & 6.85E-05 & 1.36E-04 & 1.01E-01 & 4.55E-02 & 1.12E-02 & 1.10E-02 & $\pm$0.0\\
\hline
\end{tabular}
\caption{SM Higgs branching ratios for the different $\PH \to 4\Pl$ and
$\PH \to 2\Pl2\nu$ final states and their total uncertainties (expressed in percentage). Intermediate mass range.}
\label{BR-4f1c}
\end{table}

\begin{table}[htb]
\renewcommand{\arraystretch}{1.3}
\setlength{\tabcolsep}{1.5ex}
\begin{tabular}{lccccccccc}
\hline
$\MH$  & $\PH \to 4\Pl$ & $\PH \to 4\Pl$ & $\PH \to \Pe\Pe\Pe\Pe$ & $\PH \to \Pe\Pe\mu\mu$ &
$\PH \to 2\Pl2\nu$ & $\PH \to 2\Pl2\nu$ & $\PH \to\Pe\nu\Pe \nu$ & $\PH \to\Pe\nu \mu \nu$ &$\Delta$BR  \\
%$[\UGeV]$ & $\Pl=\Pe,\mu,\tau$ & $\Pl=\Pe,\mu$ & & & $\Pl=\Pe,\mu,\tau$ - $\nu={}$any & $\Pl=\Pe,\mu$ - $\nu={}$any& & & [\%]\\ 
$[\UGeV]$ & $\Pl=\Pe,\mu,\tau$ & $\Pl=\Pe,\mu$ & & & $\Pl=\Pe,\mu,\tau$ & $\Pl=\Pe,\mu$ & & & [\%]\\ 
 & & & & & $\nu={}$any & $\nu={}$any& & & \\ 
\hline
262.0 & 3.04E-03 & 1.35E-03 & 3.38E-04 & 6.75E-04 & 8.61E-02 & 4.10E-02 & 9.57E-03 & 8.22E-03 & $\pm$0.0\\
264.0 & 3.04E-03 & 1.35E-03 & 3.38E-04 & 6.76E-04 & 8.61E-02 & 4.10E-02 & 9.57E-03 & 8.22E-03 & $\pm$0.0\\
266.0 & 3.05E-03 & 1.35E-03 & 3.39E-04 & 6.77E-04 & 8.61E-02 & 4.09E-02 & 9.56E-03 & 8.21E-03 & $\pm$0.0\\
268.0 & 3.05E-03 & 1.36E-03 & 3.39E-04 & 6.78E-04 & 8.60E-02 & 4.09E-02 & 9.56E-03 & 8.21E-03 & $\pm$0.0\\
270.0 & 3.05E-03 & 1.36E-03 & 3.40E-04 & 6.79E-04 & 8.60E-02 & 4.09E-02 & 9.56E-03 & 8.21E-03 & $\pm$0.0\\
\hline
\end{tabular}
\caption{SM Higgs branching ratios for the different $\PH \to 4\Pl$ and
$\PH \to 2\Pl2\nu$ final states and their total uncertainties (expressed in percentage). High mass range.}
\label{BR-4f1d}
\end{table}

\begin{table}[htb]
\renewcommand{\arraystretch}{1.3}
\setlength{\tabcolsep}{1.5ex}
\begin{tabular}{lccccccccc}
\hline
$\MH$  & $\PH \to 4\Pl$ & $\PH \to 4\Pl$ & $\PH \to \Pe\Pe\Pe\Pe$ & $\PH \to \Pe\Pe\mu\mu$ &
$\PH \to 2\Pl2\nu$ & $\PH \to 2\Pl2\nu$ & $\PH \to\Pe\nu\Pe \nu$ & $\PH \to\Pe\nu \mu \nu$ &$\Delta$BR  \\
%$[\UGeV]$ & $\Pl=\Pe,\mu,\tau$ & $\Pl=\Pe,\mu$ & & & $\Pl=\Pe,\mu,\tau$ - $\nu={}$any & $\Pl=\Pe,\mu$ - $\nu={}$any& & & [\%]\\ 
$[\UGeV]$ & $\Pl=\Pe,\mu,\tau$ & $\Pl=\Pe,\mu$ & & & $\Pl=\Pe,\mu,\tau$ & $\Pl=\Pe,\mu$ & & & [\%]\\ 
 & & & & & $\nu={}$any & $\nu={}$any& & & \\ 
\hline
550.0 & 2.70E-03 & 1.20E-03 & 3.00E-04 & 6.00E-04 & 6.92E-02 & 3.32E-02 & 7.69E-03 & 6.50E-03 & $\pm$2.4\\
560.0 & 2.71E-03 & 1.21E-03 & 3.01E-04 & 6.03E-04 & 6.94E-02 & 3.33E-02 & 7.71E-03 & 6.51E-03 & $\pm$2.5\\
570.0 & 2.73E-03 & 1.21E-03 & 3.03E-04 & 6.06E-04 & 6.96E-02 & 3.34E-02 & 7.74E-03 & 6.53E-03 & $\pm$2.5\\
580.0 & 2.74E-03 & 1.22E-03 & 3.04E-04 & 6.09E-04 & 6.99E-02 & 3.35E-02 & 7.76E-03 & 6.55E-03 & $\pm$2.5\\
590.0 & 2.75E-03 & 1.22E-03 & 3.06E-04 & 6.12E-04 & 7.01E-02 & 3.36E-02 & 7.79E-03 & 6.57E-03 & $\pm$2.5\\
\hline
\end{tabular}
\caption{SM Higgs branching ratios for the different $\PH \to 4\Pl$ and
$\PH \to 2\Pl2\nu$ final states and their total uncertainties (expressed in percentage). Very high mass range.}
\label{BR-4f1e}
\end{table}

\begin{table}[htb]
\renewcommand{\arraystretch}{1.3}
\setlength{\tabcolsep}{3ex}
\begin{tabular}{lccccccc}
\hline
$\MH$  & $\PH \to 2\Pl2\Pq$ & $\PH \to 2\Pl2\Pq$ & $\PH \to l\nu_l \Pq\Pq$ & $\PH \to \nu\nu \Pq\Pq$ &
$\PH \to 4\Pq$ & $\PH \to 4\Pf$ &$\Delta$BR  \\
%$[\UGeV]$& $\Pl=\Pe,\mu,\tau$ - $\Pq=\PQu\PQd\PQc\PQs\PQb$& $\Pl=\Pe,\mu$ - $\Pq=\PQu\PQd\PQc\PQs\PQb$
%&$\Pl=\Pe,\mu$ - $\Pq=\PQu\PQd\PQc\PQs\PQb$ & $\nu={}$any, $\Pq=\PQu\PQd\PQc\PQs\PQb$&  $\Pq=\PQu\PQd\PQc\PQs\PQb$&
%$f$=any  
%fermion &[\%]\\ 
$[\UGeV]$ & $\Pl=\Pe,\mu,\tau$& $\Pl=\Pe,\mu$
& $\Pl=\Pe,\mu$ & $\nu={}$any& $\Pq=\PQu\PQd\PQc\PQs\PQb$& $f={}$any
 &[\%]\\ 
& $\Pq=\PQu\PQd\PQc\PQs\PQb$ & $\Pq=\PQu\PQd\PQc\PQs\PQb$ &
$\Pq=\PQu\PQd\PQc\PQs\PQb$ & $\Pq=\PQu\PQd\PQc\PQs\PQb$ & & fermion &\\ 
\hline

90.0 & 5.87E-05 & 3.91E-05 & 3.03E-04 & 1.18E-04 & 1.05E-03 & 2.37E-03  & $\pm$5.8 \\
95.0 & 9.31E-05 & 6.21E-05 & 6.85E-04 & 1.87E-04 & 2.32E-03 & 5.17E-03  & $\pm$5.8 \\
100.0 & 1.57E-04 & 1.04E-04 & 1.61E-03 & 3.14E-04 & 5.34E-03 & 1.18E-02 & $\pm$5.7 \\
105.0 & 2.97E-04 & 1.98E-04 & 3.52E-03 & 5.95E-04 & 1.17E-02 & 2.58E-02 & $\pm$5.6 \\
110.0 & 6.06E-04 & 4.04E-04 & 6.99E-03 & 1.21E-03 & 2.34E-02 & 5.15E-02 & $\pm$5.4 \\
\hline
\end{tabular}
\caption{SM Higgs branching ratios for the different four-fermion final states 
and their total uncertainties (expressed in percentage). Very low mass range.}
\label{BR-4f2a}
\end{table}

\begin{table}[htb]
\renewcommand{\arraystretch}{1.3}
\setlength{\tabcolsep}{3ex}
\begin{tabular}{lccccccc}
\hline
$\MH$  & $\PH \to 2\Pl2\Pq$ & $\PH \to 2\Pl2\Pq$ & $\PH \to l\nu_l \Pq\Pq$ & $\PH \to \nu\nu \Pq\Pq$ &
$\PH \to 4\Pq$ & $\PH \to 4\Pf$ &$\Delta$BR  \\
%$[\UGeV]$& $\Pl=\Pe,\mu,\tau$ - $\Pq=\PQu\PQd\PQc\PQs\PQb$& $\Pl=\Pe,\mu$ - $\Pq=\PQu\PQd\PQc\PQs\PQb$ &$\Pl=\Pe,\mu$ - $\Pq=\PQu\PQd\PQc\PQs\PQb$ & $\nu={}$any, $\Pq=\PQu\PQd\PQc\PQs\PQb$&  $\Pq=\PQu\PQd\PQc\PQs\PQb$& $f={}$any fermion &[\%]\\ 
$[\UGeV]$ & $\Pl=\Pe,\mu,\tau$& $\Pl=\Pe,\mu$
& $\Pl=\Pe,\mu$ & $\nu={}$any& $\Pq=\PQu\PQd\PQc\PQs\PQb$& $f={}$any
 &[\%]\\ 
& $\Pq=\PQu\PQd\PQc\PQs\PQb$ & $\Pq=\PQu\PQd\PQc\PQs\PQb$ &
$\Pq=\PQu\PQd\PQc\PQs\PQb$ & $\Pq=\PQu\PQd\PQc\PQs\PQb$ & & fermion &\\ 
\hline
130.5 & 5.79E-03 & 3.86E-03 & 4.57E-02 & 1.15E-02 & 1.61E-01 & 3.52E-01 & $\pm$3.6 \\
131.0 & 6.00E-03 & 4.00E-03 & 4.71E-02 & 1.20E-02 & 1.66E-01 & 3.63E-01 & $\pm$3.5 \\
131.5 & 6.21E-03 & 4.14E-03 & 4.85E-02 & 1.24E-02 & 1.71E-01 & 3.74E-01 & $\pm$3.5 \\
132.0 & 6.41E-03 & 4.28E-03 & 4.99E-02 & 1.28E-02 & 1.76E-01 & 3.85E-01 & $\pm$3.4 \\
132.5 & 6.62E-03 & 4.42E-03 & 5.14E-02 & 1.32E-02 & 1.81E-01 & 3.96E-01 & $\pm$3.3 \\
\hline
\end{tabular}
\caption{SM Higgs branching ratios for the different four-fermion final states 
and their total uncertainties (expressed in percentage). Low and intermediate mass range.}
\label{BR-4f2b}
\end{table}

\begin{table}[htb]
\renewcommand{\arraystretch}{1.3}
\setlength{\tabcolsep}{3ex}
\begin{tabular}{lccccccc}
\hline
$\MH$  & $\PH \to 2\Pl2\Pq$ & $\PH \to 2\Pl2\Pq$ & $\PH \to l\nu_l \Pq\Pq$ & $\PH \to \nu\nu \Pq\Pq$ &
$\PH \to 4\Pq$ & $\PH \to 4\Pf$ &$\Delta$BR  \\
%$[\UGeV]$& $\Pl=\Pe,\mu,\tau$ - $\Pq=\PQu\PQd\PQc\PQs\PQb$& $\Pl=\Pe,\mu$ - $\Pq=\PQu\PQd\PQc\PQs\PQb$ &$\Pl=\Pe,\mu$ - $\Pq=\PQu\PQd\PQc\PQs\PQb$ & $\nu={}$any, $\Pq=\PQu\PQd\PQc\PQs\PQb$&  $\Pq=\PQu\PQd\PQc\PQs\PQb$& $f={}$any fermion &[\%]\\ 
$[\UGeV]$ & $\Pl=\Pe,\mu,\tau$& $\Pl=\Pe,\mu$
& $\Pl=\Pe,\mu$ & $\nu={}$any& $\Pq=\PQu\PQd\PQc\PQs\PQb$& $f={}$any
 &[\%]\\ 
& $\Pq=\PQu\PQd\PQc\PQs\PQb$ & $\Pq=\PQu\PQd\PQc\PQs\PQb$ &
$\Pq=\PQu\PQd\PQc\PQs\PQb$ & $\Pq=\PQu\PQd\PQc\PQs\PQb$ & & fermion &\\ 
\hline
172.0 & 3.67E-03 & 2.44E-03 & 1.41E-01 & 7.30E-03 & 4.50E-01 & 9.89E-01 & $\pm$0.0 \\
174.0 & 4.19E-03 & 2.79E-03 & 1.41E-01 & 8.34E-03 & 4.51E-01 & 9.90E-01 & $\pm$0.0 \\
176.0 & 4.98E-03 & 3.32E-03 & 1.40E-01 & 9.92E-03 & 4.51E-01 & 9.91E-01 & $\pm$0.0 \\
178.0 & 6.24E-03 & 4.16E-03 & 1.39E-01 & 1.24E-02 & 4.52E-01 & 9.91E-01 & $\pm$0.0 \\
180.0 & 8.47E-03 & 5.65E-03 & 1.37E-01 & 1.69E-02 & 4.53E-01 & 9.92E-01 & $\pm$0.0 \\
\hline
\end{tabular}
\caption{SM Higgs branching ratios for the different four-fermion final states and their total uncertainties (expressed in percentage). Intermediate mass range.}
\label{BR-4f2c}
\end{table}

\begin{table}[htb]
\renewcommand{\arraystretch}{1.3}
\setlength{\tabcolsep}{3ex}
\begin{tabular}{lccccccc}
\hline
$\MH$  & $\PH \to 2\Pl2\Pq$ & $\PH \to 2\Pl2\Pq$ & $\PH \to l\nu_l \Pq\Pq$ & $\PH \to \nu\nu \Pq\Pq$ &
$\PH \to 4\Pq$ & $\PH \to 4\Pf$ &$\Delta$BR  \\
%$[\UGeV]$& $\Pl=\Pe,\mu,\tau$ - $\Pq=\PQu\PQd\PQc\PQs\PQb$& $\Pl=\Pe,\mu$ - $\Pq=\PQu\PQd\PQc\PQs\PQb$ &$\Pl=\Pe,\mu$ - $\Pq=\PQu\PQd\PQc\PQs\PQb$ & $\nu={}$any, $\Pq=\PQu\PQd\PQc\PQs\PQb$&  $\Pq=\PQu\PQd\PQc\PQs\PQb$& $f={}$any fermion &[\%]\\ 
$[\UGeV]$ & $\Pl=\Pe,\mu,\tau$& $\Pl=\Pe,\mu$
& $\Pl=\Pe,\mu$ & $\nu={}$any& $\Pq=\PQu\PQd\PQc\PQs\PQb$& $f={}$any
 &[\%]\\ 
& $\Pq=\PQu\PQd\PQc\PQs\PQb$ & $\Pq=\PQu\PQd\PQc\PQs\PQb$ &
$\Pq=\PQu\PQd\PQc\PQs\PQb$ & $\Pq=\PQu\PQd\PQc\PQs\PQb$ & & fermion &\\ 
\hline
262.0 & 4.22E-02 & 2.81E-02 & 1.02E-01 & 8.41E-02 & 4.64E-01 & 9.98E-01 & $\pm$0.0 \\
264.0 & 4.23E-02 & 2.82E-02 & 1.02E-01 & 8.42E-02 & 4.64E-01 & 9.98E-01 & $\pm$0.0 \\
266.0 & 4.23E-02 & 2.82E-02 & 1.02E-01 & 8.43E-02 & 4.64E-01 & 9.98E-01 & $\pm$0.0 \\
268.0 & 4.24E-02 & 2.83E-02 & 1.02E-01 & 8.44E-02 & 4.64E-01 & 9.98E-01 & $\pm$0.0 \\
270.0 & 4.24E-02 & 2.83E-02 & 1.02E-01 & 8.45E-02 & 4.64E-01 & 9.98E-01 & $\pm$0.0 \\
\hline
\end{tabular}
\caption{SM Higgs branching ratios for the different four-fermion final states and their total uncertainties (expressed in percentage). High mass range.}
\label{BR-4f2d}
\end{table}

\begin{table}[htb]
\renewcommand{\arraystretch}{1.3}
\setlength{\tabcolsep}{3ex}
\begin{tabular}{lccccccc}
\hline
$\MH$  & $\PH \to 2\Pl2\Pq$ & $\PH \to 2\Pl2\Pq$ & $\PH \to l\nu_l \Pq\Pq$ & $\PH \to \nu\nu \Pq\Pq$ &
$\PH \to 4\Pq$ & $\PH \to 4\Pf$ &$\Delta$BR  \\
%$[\UGeV]$& $\Pl=\Pe,\mu,\tau$ - $\Pq=\PQu\PQd\PQc\PQs\PQb$& $\Pl=\Pe,\mu$ - $\Pq=\PQu\PQd\PQc\PQs\PQb$
%&$\Pl=\Pe,\mu$ - $\Pq=\PQu\PQd\PQc\PQs\PQb$ &, $\Pq=\PQu\PQd\PQc\PQs\PQb$&
%$\Pq=\PQu\PQd\PQc\PQs\PQb$& $f={}$any  fermion &[\%]\\ 
$[\UGeV]$ & $\Pl=\Pe,\mu,\tau$& $\Pl=\Pe,\mu$
& $\Pl=\Pe,\mu$ & $\nu={}$any& $\Pq=\PQu\PQd\PQc\PQs\PQb$& $f={}$any
 &[\%]\\ 
& $\Pq=\PQu\PQd\PQc\PQs\PQb$ & $\Pq=\PQu\PQd\PQc\PQs\PQb$ &
$\Pq=\PQu\PQd\PQc\PQs\PQb$ & $\Pq=\PQu\PQd\PQc\PQs\PQb$ & & fermion &\\ 
\hline
550.0 & 3.75E-02 & 2.50E-02 & 8.06E-02 & 7.45E-02 & 3.80E-01 & 8.16E-01 & $\pm$2.4 \\
560.0 & 3.76E-02 & 2.51E-02 & 8.09E-02 & 7.49E-02 & 3.81E-01 & 8.19E-01 & $\pm$2.5 \\
570.0 & 3.78E-02 & 2.52E-02 & 8.11E-02 & 7.52E-02 & 3.82E-01 & 8.22E-01 & $\pm$2.5 \\
580.0 & 3.80E-02 & 2.53E-02 & 8.13E-02 & 7.56E-02 & 3.84E-01 & 8.25E-01 & $\pm$2.5 \\
590.0 & 3.82E-02 & 2.54E-02 & 8.16E-02 & 7.60E-02 & 3.85E-01 & 8.28E-01 & $\pm$2.5 \\
\hline
\end{tabular}
\caption{SM Higgs branching ratios for the different four-fermion final states and their total uncertainties (expressed in percentage). Very high mass range.}
\label{BR-4f2e}
\end{table}

%%% Local Variables: 
%%% mode: latex
%%% TeX-master: ../LHCHiggsBRNote.tex
%%% TeX-master: "LHCHiggsBRNote"
%%% TeX-master: t
%%% End: 